\documentclass[reprint,twocolumn,longbibliography,superscriptaddress,
showpacs,preprintnumbers,
notitlepage,amsmath,amssymb,
prl,
]{revtex4-1}

\usepackage{graphicx}
\usepackage{dcolumn}
\usepackage{bm}
\usepackage{bbm}
\usepackage{diagbox}

\usepackage{graphicx}
\usepackage{enumerate}
\usepackage{subfigure}
\usepackage{epstopdf}
\usepackage[colorlinks,
linkcolor=red,
anchorcolor=black,
citecolor=blue]{hyperref}
\usepackage{amssymb}
\usepackage{framed}
\usepackage{tabularx}
\usepackage{booktabs}
\usepackage{float}
\usepackage[table]{xcolor}
\usepackage{array}
\usepackage{tikz}
\hypersetup{colorlinks=true}

\begin{document}
\title{Strongly correlated crystalline higher-order topological phases in two-dimensional systems: A coupled-wire study
}
\author{Jian-Hao Zhang}
\email{jianhaozhang11@psu.edu}
\affiliation{Department of Physics, The Chinese University of Hong Kong, Shatin, New Territories, Hong Kong, China}
\affiliation{Department of Physics, The Pennsylvania State University, University Park, Pennsylvania 16802, USA}

\newcommand{\lra}{\longrightarrow}
\newcommand{\xra}{\xrightarrow}
\newcommand{\ra}{\rightarrow}
\newcommand{\bs}{\boldsymbol}
\newcommand{\ul}{\underline}
\newcommand{\1}{\text{\uppercase\expandafter{\romannumeral1}}}
\newcommand{\2}{\text{\uppercase\expandafter{\romannumeral2}}}
\newcommand{\3}{\text{\uppercase\expandafter{\romannumeral3}}}
\newcommand{\4}{\text{\uppercase\expandafter{\romannumeral4}}}
\newcommand{\5}{\text{\uppercase\expandafter{\romannumeral5}}}
\newcommand{\6}{\text{\uppercase\expandafter{\romannumeral6}}}

\begin{abstract}
Coupled-wire constructions have been widely applied to quantum Hall systems and symmetry-protected topological (SPT) phases. In this Letter, we use the coupled one-dimensional nonchiral Luttinger liquids with domain-wall structured mass terms as quantum wires to construct the crystalline higher-order topological superconductors (HOTSC) in two-dimensional interacting fermionic systems by two representative examples: $D_4$-symmetric class-D HOTSC and $C_4$-symmetric class-BD\1 HOTSC, with Majorana corner modes on the edge. Furthermore, based on the coupled-wire constructions, the quantum phase transition between different phases of 2D HOTSC by tuning the inter-wire coupling are investigated in a straightforward way.
\end{abstract}

\maketitle


\textit{Introduction} -- Topological phases of quantum matter have become one of the greatest triumph of condensed matter physics since the discovery of fractional quantum Hall effect \cite{FQHE,Laughlin}. Topological order defined by patterns of long-range entanglement provides a systematic way of understanding topological phases of quantum matter \cite{entanglement}. Furthermore, the interplay between symmetry and topology plays a central role in the topological phases of quantum matter. In particular, symmetry-protected topological (SPT) phases has been systematically constructed and classified in short-range entangled systems \cite{ZCGu2009,chen11a,XieChenScience,cohomology,Senthil_2015,E8,Lu12,invertible2,invertible3,special,general1,general2,Kapustin2014,Kapustin2015,Kapustin2017,Gu-Levin,gauging1,gauging3,dimensionalreduction,gauging2,2DFSPT,braiding,Ning21a}. An elegant example of SPT phases is topological insulator, protected by time-reversal and charge-conservation symmetry~\cite{KaneRMP,ZhangRMP}. Recently, crystalline SPT phases have been intensively studied \cite{TCI,Fu2012,ITCI,reduction,building,correspondence,SET,230,BCSPT,Jiang2017,Kane2017,Shiozaki2018,ZDSong2018,defect,realspace,KenX,rotation,dihedral,LuX,YMLu2018,Cheng2018,Hermele2018,Po2020,Huang2020PRR,Huang2021PRR,wallpaper,PEPS,Maissam2020,Maissam2021,Max18,JosephMeng19,3Dpoint}, with great opportunities for experimental realizations~\cite{TCIrealization1,TCIrealization2,TCIrealization3,TCIrealization4}. In particular, different from internal SPT phases, the boundaries of 2D crystalline SPT phases are almost gapped but with protected 0D \textit{corner} zero modes. This type of topological phases are called \textit{higher-order topological phases}  \cite{Wang2018,Yan2018,Nori2018,Wangyuxuan2018,Ryu2018,Zhang2019,Hsu2018,bultinck2019three,Roy_2020,Roy_2021,Laubscher_2019,Laubscher_2020}. 

The study of higher-order topological phases mainly focus on free-fermion systems, because interactions and crystalline symmetries restrict the analytical study of lattice model, only numerics on finite size lattice can give some insights. On the other hand, a clear and powerful tool of studying topological phases of quantum matter is coupled-wire construction \cite{Karlhede2008,Lubensky2002,Kane2014,Lecheminant2017,Thomale2015,Sagi_2015,Seroussi_2014,Iadecola_2016,Neupert_2014,Klinovaja_2014,Klinovaja_2015,Klinovaja_2014a}. One decomposes a higher-dimensional system into an assembly of 1D quantum wires, and topological properties then arise from the suitable couplings of them. A unique advantage of coupled-wire construction is that different from higher-dimensional quantum field theory, the powerful bosonization technique of one-dimensional subsystems can be used to challenge the strong interaction effects. Different phases are manifested by patterns of coupled wires and the quantum phase transition of different phases is controlled by tuning inter-wire couplings directly. Therefore, an important open question arises: if the strongly correlated higher-order topological phases can be constructed by coupled-wire perspective?

In this Letter, we systematically construct the crystalline HOTSC in two-dimensional interacting fermionic systems by coupling the circular 1D nonchiral Luttinger liquids with domain-wall structured mass terms as quantum wires, via two typical intriguing interacting examples: $D_4$-symmetric class-D HOTSC and $C_4$-symmetric class-BD\1 HOTSC, whose higher-order edge modes are Majorana zero modes (MZMs) \cite{rotation,dihedral}. By suitable inter-wire tunneling/interaction, several 1D quantum wires are assembled and fully gapped, living few dangling quantum wires at the edge or near the center of the systems. Near the center, the dangling quantum wires are fully gapped by intra-wire interactions; on the edge, the dangling quantum wires explicitly manifest the higher-order topological edge modes of 2D HOTSC by their domain-wall structure. Different 2D HOTSCs are characterized by different patterns of coupled-wire. Lattice translation symmetry can also be imposed straightforwardly. Furthermore, with concrete coupled-wire construction of 2D HOTSC, we directly investigate the quantum phase transitions by tuning different coupling constants of inter-wire interactions. We stress that our arguments are not sensitive to specific geometry of quantum wires: the calculations are applicable to any geometry respecting the specific crystalline symmetry, we choose circular geometry for calculational convenience.

\textit{$D_4$-symmetric class-D HOTSC} -- For 2D $D_4$-symmetric systems with spinless fermions, there is an intriguing interacting 2D HOTSC with protected Majorana corner modes $\xi_{k}$ and $\xi_{k}'$ ($k=1,2,3,4$) that can be reformulated to complex fermions $c_{k}^\dag=(\xi_{k}+i\xi_{k}')/\sqrt{2}$ (see Fig. \ref{CWC}). In this section we construct this phase by an ``almost free'' coupled-wires, with necessary interaction only defined near the $D_4$-center. These Majorana corner modes are also reformulated in terms of domain walls of 1D nonchiral Luttinger liquids \cite{BBC}. Consider $2n$ decoupled 1D quantum wires with circular geometry (see Fig. \ref{CWC}), the Lagrangian of $j^{\mathrm{th}}$ quantum wire is:
\begin{align}
\mathcal{L}_{0}^j=\frac{K_{IJ}^j}{4\pi}\left(\partial_\theta\phi_I^j\right)\left(\partial_t\phi_J^j\right)+\frac{V_{IJ}^j}{8\pi}\left(\partial_\theta\phi_I^j\right)\left(\partial_\theta\phi_J^j\right)
\label{Luttinger}
\end{align}
where $\phi^j=(\phi_1^j,\phi_2^j)^T$ is the 2-component bosonic field of $j^{\mathrm{th}}$ quantum wire and $K^j=\sigma^z$ as the $K$-matrix of the topological term \cite{supplementary}. The total Lagrangian of decoupled wires is: $\mathcal{L}_0=\sum_{j=1}^{2n}\mathcal{L}_0^j$. The $D_4$ symmetry properties of these bosonic fields are ($\bs{R}\in C_4/\bs{M}_1\in\mathbb{Z}_2^M$ is rotation/reflection generator of $D_4=C_4\rtimes\mathbb{Z}_2^M$ symmetry):
\begin{align}
\begin{aligned}
\bs{R}:&\left\{
\begin{aligned}
&\phi_1^j(\theta)\mapsto-\phi_1^j(\theta+\pi/2)\\
&\phi_2^j(\theta)\mapsto-\phi_2^j(\theta+\pi/2)+\pi
\end{aligned}
\right.\\
\bs{M}_1:&\left\{
\begin{aligned}
&\phi_1^j(\theta)\mapsto-\phi^j_2(2\pi-\theta)+\pi/2\\
&\phi_2^j(\theta)\mapsto-\phi^j_1(2\pi-\theta)+\pi/2
\end{aligned}
\right.
\end{aligned}
\label{D4 symmetry}
\end{align}
\begin{figure}
\includegraphics[width=0.48\textwidth]{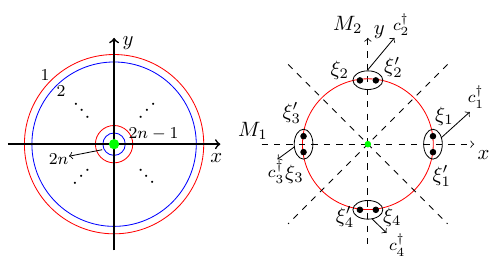}
\caption{Coupled-wire construction of 2D fermionic crystalline HOTSC. Right panel: dangling gapless modes of $D_4$-symmetric class-D or $C_4$-symmetric class-BD\1 HOTSC. Dashed lines are reflection axes for $D_4$ symmetry.}
\label{CWC}
\end{figure}
To figure out the Majorana corner modes of $D_4$-symmetric HOTSC, we should further introduce the mass term with domain wall structure of each quantum wire:
\begin{align}
\mathcal{L}_{\mathrm{wall}}^j=m\sin(2\theta)\cdot\cos\left[\phi_1^j(\theta)+\phi_2^j(\theta)\right]
\end{align}
where $\mathcal{L}_{\mathrm{wall}}^j$ is symmetric under (\ref{D4 symmetry}), and $\mathcal{L}_{\mathrm{wall}}=\sum_{j=1}^{2n}\mathcal{L}_{\mathrm{wall}}^j$. For each quantum wire with a domain-wall structured mass term, there are four complex fermion zero modes at poles $c_{1,2,3,4}^\dag$ of the circle, with $\theta=0,\pi/2,\pi,3\pi/2$ because of the vanishing mass term, which are equivalent to eight MZMs. These dangling 0D gapless modes cannot be gapped in a $D_4$-symmetric way. 

Subsequently we define two types of $D_4$-symmetric (\ref{D4 symmetry}) inter-wire tunneling that couple the $(2j-2+k)^{\mathrm{th}}$ and $(2j-1+k)^{\mathrm{th}}$ quantum wires ($m_1,m_2<m$, $k=1,2$):
\begin{align}
\mathcal{L}_{ck}^j=m_k\sum\limits_{\alpha=1}^2\cos\left[\phi_\alpha^{2j-2+k}(\theta)-\phi_\alpha^{2j-1+k}(\theta)\right]
\end{align}
and $\mathcal{L}_{ck}=\sum_{j=1}^n\mathcal{L}_{ck}^j$. There are two extreme cases: $m_1\ne0,m_2=0/m_1=0,m_2\ne0$ corresponds to the phase that $\mathcal{L}_{c1}/\mathcal{L}_{c2}$ dominates the inter-wire physics. For $\mathcal{L}_{c1}$-dominant phase, the $(2j-1)^{\mathrm{th}}$ and $2j^{\mathrm{th}}$ wires are paired up and gapped, hence the corresponding system is fully gapped on a open circle and topological trivial. 

\begin{figure}
\includegraphics[width=0.45\textwidth]{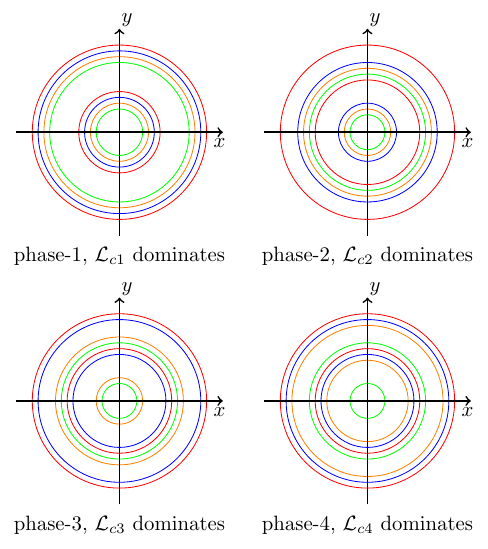}
\caption{4 distinct phases of 2D BD\1-class $(C_4\times\mathbb{Z}_2^T)$-symmetric system. Each four 1D wires with narrower intervals can be gapped by inter-wire interactions $\mathcal{L}_{ck}$, and the 1D wire near $C_4$-center can be gapped by intra-wire interactions $\mathcal{L}_{\mathrm{int}}$.}
\label{BDI phase}
\end{figure}

For $\mathcal{L}_{c2}$-dominant phase, the $2j^{\mathrm{th}}$ and $(2j+1)^{\mathrm{th}}$ wires are paired up and gapped, hence all 1D wires except $1^{\mathrm{st}}$ and $2n^{\mathrm{th}}$ are gapped. The $1^{\mathrm{st}}$ wire on the edge of the system presents 4 complex fermions/8 MZMs at poles of circle, which are exactly the second-order topological surface modes of 2D $D_4$-symmetric class-D HOTSC. Near the $D_4$-center, there are also gapless modes on the $2n^{\mathrm{th}}$ quantum wire. Distinct from quantum wires away from $D_4$-center, bosonic field $\phi^{2n}$ of the $2n^{\mathrm{th}}$ quantum wires with different polar angles can tunnel to/interact with the field at other places. Consider two interacting terms of $2n^{\mathrm{th}}$ quantum wire near the $D_4$-center:
\begin{align}
\mathcal{L}_{\mathrm{int}}=m'\sum\limits_{\beta=1}^2&\cos\left(\sum\limits_{\alpha=1}^2\left[\phi_\alpha^{2n}(\theta)-\phi_\alpha^{2n}(\beta\pi-\theta)\right]\right)
\end{align}
i.e., the intra-wire couplings of the $2n^{\mathrm{th}}$ wire lead to a fully gapped bulks. Equivalently, a nontrivial 2D class-D $D_4$-symmetric HOTSC are described by 1D coupled wires with Lagrangian $\mathcal{L}_{D_4}^D=\mathcal{L}_0+\mathcal{L}_{\mathrm{wall}}+\mathcal{L}_{c2}+\mathcal{L}_{\mathrm{int}}$. The intriguing interacting nature of this HOTSC is reflected by $\mathcal{L}_{\mathrm{int}}$ near the $D_4$-center. On the other hand, the physics away from the $D_4$-center is well-understood on the noninteracting level. The classification of 2D class-D HOTSC is $\mathbb{Z}_2$, composed by phases dominated by inter-wire coupling $\mathcal{L}_{c1}$ and $\mathcal{L}_{c2}$ (see Fig. \ref{QPT}).

\textit{$C_4$-symmetric class-BD\1 HOTSC} -- For 2D BD\1-class systems with $C_4$-symmetry, there is another type of intriguing interacting 2D HOTSCs \cite{rotation}. We construct these phases by ``interacting'' coupled-wires in this section. Consider $4n$ 1D circular quantum wires, each wire is described by Lagrangian (\ref{Luttinger}) with $\phi^j=(\phi_1^j,\phi_2^j,\phi_3^j,\phi_4^j)^T$ as 4-component bosonic field of $j^{\mathrm{th}}$ quantum wire,  $K^j=\sigma^z\oplus\sigma^z$ as the topological $K$-matrix, and the total Lagrangian of all $4n$ 1D quantum wires is $\mathcal{L}_0=\sum_{j=1}^{4n}\mathcal{L}_0^j$. The $(C_4\times\mathbb{Z}_2^T)$-symmetry properties are defined as \cite{supplementary}:
\begin{align}
\bs{R}:\left\{
\begin{aligned}
&\phi_1^j\mapsto\phi_1^j+\phi_3^j-\phi_4^j-\pi/2\\
&\phi_2^j\mapsto\phi_2^j-\phi_3^j+\phi_4^j+\pi/2\\
&\phi_3^j\mapsto\phi_1^j+\phi_2^j-\phi_3^j+\pi/2\\
&\phi_4^j\mapsto\phi_1^j+\phi_2^j-\phi_4^j+\pi/2
\end{aligned}
\right.~,~~\theta\mapsto\theta+\pi/2
\label{BDI-R}
\end{align}
\begin{align}
\mathcal{T}:\left\{
\begin{aligned}
&\phi_1^j(\theta)\mapsto\phi_2^j(\theta)-\phi_3^j(\theta)+\phi_4^j(\theta)\\
&\phi_2^j(\theta)\mapsto\phi_1^j(\theta)+\phi_3^j(\theta)-\phi_4^j(\theta)+\pi\\
&\phi_3^j(\theta)\mapsto\phi_1^j(\theta)+\phi_2^j(\theta)-\phi_4^j(\theta)+\pi\\
&\phi_4^j(\theta)\mapsto\phi_1^j(\theta)+\phi_2^j(\theta)-\phi_3^j(\theta)
\end{aligned}
\right.
\label{BDI-T}
\end{align}

Then we repeatedly figure out the Majorana corner modes by backscattering terms with domain-wall structure: for each quantum wire, we introduce a symmetric [cf. Eqs. (\ref{BDI-R}) and (\ref{BDI-T})] mass term:
\begin{align}
\mathcal{L}_{\mathrm{wall}}^j=&m\sum\limits_{\alpha=1}^2\cos\left(\theta-\frac{\alpha\pi}{2}\right)\cdot\cos\left[\phi_\alpha^j(\theta)-\phi_{5-\alpha}^j(\theta)\right]
\label{BDI domain wall}
\end{align}
And $\mathcal{L}_{\mathrm{wall}}=\sum_{j=1}^{4n}\mathcal{L}_{\mathrm{wall}}^j$. For each quantum wire, there are 4 gapless comlex fermions $c_{k}^\dag$ (8 MZMs $\xi_{k}$ and $\xi_{k}'$) at poles of the circle, two of them at north and south poles are from the first term of (\ref{BDI domain wall}) and other two at east and west poles are from the second term of (\ref{BDI domain wall}). These dangling gapless modes cannot be gapped in a $(C_4\times\mathbb{Z}_2^T)$-symmetric way. 

\begin{figure}
\includegraphics[width=0.47\textwidth]{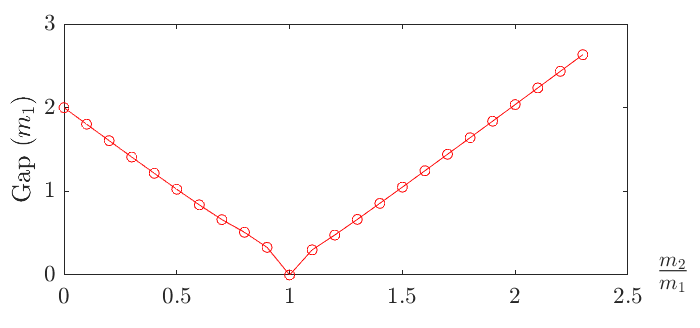}
\caption{Bulk gap of 2D $D_4$-symmetric Lagrangian $\mathcal{L}_0+\mathcal{L}_{c1}+\mathcal{L}_{c2}+\mathcal{L}_{\mathrm{int}}$, with respect to the ratio $m_2/m_1$.}
\label{gap}
\end{figure}

Subsequently we consider $(C_4\times\mathbb{Z}_2^T)$-symmetric [cf. Eqs. (\ref{BDI-R}) and (\ref{BDI-T})] inter-wire interactions including four 1D quantum wires ($k=1,2,3,4$) \cite{supplementary}:
\begin{widetext}
\begin{align}
\mathcal{L}_{ck}^j=m_k\sum\limits_{\alpha=1}^2&\cos\left[\phi_\alpha^{4j-4+k}-\phi_{5-\alpha}^{4j-4+k}+\phi_\alpha^{4j-3+k}-\phi_{5-\alpha}^{4j-3+k}\right]+\cos\left[\phi_\alpha^{4j-4+k}-\phi_{5-\alpha}^{4j-4+k}+\phi_\alpha^{4j-2+k}-\phi_{5-\alpha}^{4j-2+k}\right]\nonumber\\
+&\cos\left[\phi_\alpha^{4j-2+k}-\phi_{5-\alpha}^{4j-2+k}+\phi_\alpha^{4j-1+k}-\phi_{5-\alpha}^{4j-1+k}\right]+\cos\left[\phi_\alpha^{4j-3+k}-\phi_{5-\alpha}^{4j-3+k}+\phi_\alpha^{4j-1+k}-\phi_{5-\alpha}^{4j-1+k}\right]
\end{align}
\end{widetext}
and the total Lagrangian of inter-wire couplings is $\mathcal{L}_{ck}=\sum_{j=1}^{n-1}\mathcal{L}_{ck}^j$. There are four extreme cases: $m_k\ne0$ ($k=1,2,3,4$) as the only nonzero index in $m_{1,2,3,4}$, which corresponds to the phase that $\mathcal{L}_{ck}$ deminates the inter-wire physics. For $\mathcal{L}_{c1}$-dominant phase, the $(4j-k)^{\mathrm{th}}$ ($k=0,1,2,3$) quantum wires are assembled and gapped, hence the spectrum is fully gapped on a 2D open circle, and the corresponding phase is topological trivial. 

For $\mathcal{L}_{c4}$-dominant phase with $m_4\ne0$ and $m_{1,2,3}=0$, by applying $\mathcal{L}_{\mathrm{wall}}$ and $\mathcal{L}_{c4}$, the $(4j+3-k)^{\mathrm{th}}$ quantum wires are assembled and gapped, and there are only 4 quantum wires remain gapless: $1^{\mathrm{st}}$, $2^{\mathrm{nd}}$, $3^{\mathrm{rd}}$ on the edge, and $4n^{\mathrm{th}}$ near the $C_4$-center. On the edge, $1^{\mathrm{st}}$, $2^{\mathrm{nd}}$ and $3^{\mathrm{rd}}$ quantum wires with dangling gapless modes are treated as the higher-order edge state of 2D $C_4$-symmetric class-BD\1 HOTSC; near the $C_4$-center, in order to obtain a HOTSC, we should further add some intra-wire interactions to fully gap the $4n^{\mathrm{th}}$ quantum wire in order to get a fully-gapped bulk state. Consider the 4-body interacting terms of $4n^{\mathrm{th}}$ quantum wire, composed by the backscatterings of bosonic fields $\phi_{1,2,3,4}^{4n}$ with different polar angles \cite{supplementary}:
\begin{align}
\mathcal{L}_{\mathrm{int}}=m'\sum\limits_{\alpha,\beta=1}^2\cos&\left[\phi_{\alpha}^{4n}(\theta)-\phi_{5-\alpha}^{4n}(\theta)+\phi_{\alpha}^{4n}(\theta+\beta\pi/2)\right.\nonumber\\
&\left.-\phi_{5-\alpha}^{4n}(\theta+\beta\pi/2)\right]
\end{align}
i.e., the intra-wire interactions of the $4n^{\mathrm{th}}$ quantum wire lead to a fully gapped bulk, and a nontrivial 2D $(C_4\times\mathbb{Z}_2^T)$-symmetric HOTSC with spinless fermions are described by 1D coupled quantum wires with Lagrangian $\mathcal{L}_{C_4}^{\mathrm{BD\1}}=\mathcal{L}_0+\mathcal{L}_{\mathrm{wall}}+\mathcal{L}_{c4}+\mathcal{L}_{\mathrm{int}}$. Similar for $\mathcal{L}_{c2}$ and $\mathcal{L}_{c3}$ dominant phases, and there are 4 topological distinct phases for 2D BD\1-class $(C_4\times\mathbb{Z}_2^T)$-symmetric system, see Fig. \ref{BDI phase}. The interacting nature of these topological phases are reflected by inter-wire interactions $\mathcal{L}_{ck}$ and intra-wire interactions near the $C_4$-center, $\mathcal{L}_{\mathrm{int}}$. The classification of 2D class-BD\1 HOTSC is $\mathbb{Z}_4$, composed by phases dominated by inter-wire couplings $\mathcal{L}_{ck}$ ($k=1,2,3,4$).

The coupled-wire construction is not limited to superconductors, it is also applicable to topological insulators: the only difference is that the Luttinger liquid (\ref{Luttinger}) should respect the $U(1)$ charge conservation.

\begin{figure}
\includegraphics[width=0.48\textwidth]{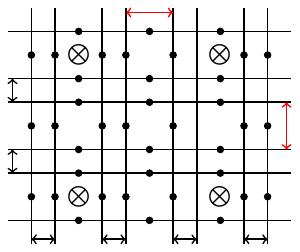}
\caption{Coupled wires of 2D $p4mm$-symmetric class-D HOTSC. Each ``$\bigotimes$'' symbol depicts a center of $D_4$, each line (horizontal or verticle) depicts a quantum wire, and each dot depicts a domain wall described by $\mathcal{L}_{\mathrm{wall}}^j$. Red and black double arrows depict different inter-wire couplings $\mathcal{L}_{c1}$ and $\mathcal{L}_{c2}$.}
\label{translation}
\end{figure}

\textit{Imposition of Lattice Translation} -- With point group symmetric cases, by $D_4$-symmetric class-D example, we demonstrate that the imposition of lattice translation symmetry is straightforward: impose the lattice translation to $D_4$ leads to $p4mm$ wallpaper group. We arrange 8 quantum wires near each $D_4$-center (4 vertical and 4 horizontal, see Fig. \ref{translation}), different topological phases are also controlled by patterns of inter-wire couplings: topological trivial phase is dominated by $\mathcal{L}_{c1}$ (black double arrows in Fig. \ref{translation}), and nontrivial phase is deminated by $\mathcal{L}_{c2}$ (red double arrows in Fig. \ref{translation}) and $\mathcal{L}_{\mathrm{int}}$ at each $D_4$-center in order to the fully gapped bulk. 

\textit{Quantum phase transition of HOTSC} -- Coupled-wire picture serves a unique platform for investigating the quantum phase transition (QPT) of 2D HOTSC because of its clear formulations. In this section, we elucidate the QPT of 2D intriguing interacting $D_4$-symmetric HOTSC as a representative example. Consider the $D_4$-symmetric Lagrangian $\mathcal{L}_0+\mathcal{L}_{c1}+\mathcal{L}_{c2}+\mathcal{L}_{\mathrm{int}}$, above we have discussed two extreme cases with $m_1=0/m_2=0$, derive two distinct phases characterized by appearence of Majorana corner modes on the edge ($1^{\mathrm{st}}$ quantum wire). Now we suppose $m=10m_1$ and set both $m_1$ and $m_2$ finite and study the possible QPT by tuning their ratio $m_2/m_1$. As summarized in Fig. \ref{gap}, turn on $m_2$ in $m_2<m_1$ regime, the system remains fully gapped with narrower gap; at $m_1=m_2$, the gap closes and the system becomes critical; keep increasing $m_2$ toward $m_2>m_1$ regime, the system reopens a \textit{bulk} gap but leaving several gapless modes on the edge, which are exactly the Majorana domain walls of 1D quantum wire on the edge. Therefore, we conclude that there is a clear quantum phase transition from trivial state to 2D $D_4$-symmetric HOTSC at $m_1=m_2$ point. Equivalently, this quantum phase transition is characterized by different inter-wire entanglement patterns of 1D quantum wires, as illustrated in Fig. \ref{QPT}.

\begin{figure}
\includegraphics[width=0.48\textwidth]{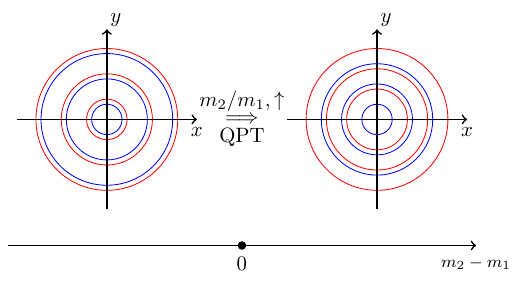}
\caption{Quantum phase transition of 2D $D_4$-symmetric HOTSC, with respect to $m_2-m_1$. Each pair of quantum wires with narrower intervals are coupled.}
\label{QPT}
\end{figure}

For 2D $C_4$-symmetric class-BD\1 system, the quantum phase transitions can be described in a similar way with little complications. For this case, there are four distinct phases controlled by four different parameters $m_{1,2,3,4}$ (see Fig. \ref{BDI phase}). As an example, for the quantum phase transition between phase-2 and phase-3, we set $m_1=m_4=0$ and investigate the bulk gap by tuning the ratio $m_2/m_3$. Heuristically, we see that the system will be critical for $m_2=m_3$, hence there will be a quantum phase transition at this point \cite{supplementary}. As a matter of fact, distinct phases of 2D HOTSC are controlled by different patterns of inter-wire entanglements, and their quantum phase transitions can be manipulated by tuning the intensities of different types of inter-wire couplings. In other words, coupled-wire construction provides a straightforward way of comprehending the quantum phase transitions of 2D HOTSCs, by tuning the inter-wire couplings to control the patterns of inter-wire entanglement. Different phases are manifested by different numbers of Majorana zero modes at each pole of the outer-most quantum wire.

\textit{Experimental Implications} -- In this Letter, the explicit manifestations of second-order modes on the edge of the systems by domain-wall structured quantum wires serves a direct opportunity for observing the higher-order topological phases by tunneling spectroscopic measurements. Recently, the coupled-wire picture is straightforwardly manifested in two-dimensional Moir\'e superlattices \cite{Wu_2019,Wang_2022}. In particular, in Ref. \cite{Wang_2022}, one-dimensional Luttinger liquids behavior has been explicitly observed in 2D bilayer WTe$_2$ Moir\'e superlattice by direct transport measurements. Hence our approach can directly be applied to Moir\'e superlattice. 

\textit{Conclusion and Discussion} -- Coupled-wire construction is a celebrated aspect in topological phases of quantum matter, for both long-range and short-range entangled systems. In this Letter, we establish the coupled-wire construction of 2D intriguing interacting fermionic crystalline HOTSC, with two representative examples: 2D $D_4$-symmetric class-D and $C_4$-symmetric class-BD\1 HOTSC phases. An indispensable advantage of coupled-wire construction is that the powerful bosonization technique can be utilized, and the inter-wire couplings can be straightforwardly involved by many-body backscattering terms in the Lagrangian. With this advantage, we use the 1D nonchiral Luttinger liquid with a domain-wall structured mass term as an ``almost gapped'' 1D quantum wire. Based on these quantum wires, we introduce some suitable inter-wire couplings in order to gap out the bulk by assemblies of quantum wires. The remaining ungapped quantum wires on the edge are treated as the edge theory of 2D HOTSC. Near the center of point group, the ungapped quantum wires are gapped by interactions of bosonic fields at different places. The lattice translation symmetry can be straightforwardly imposed. Distinct HOTSCs are manifested by different patterns of inter-wire entanglement. Furthermore, the concrete coupled-wire constructions serve a straightforward way to comprehend the quantum phase transitions of 2D HOTSCs, by directly tuning the inter-wire couplings to control the inter-wire entanglement patterns. The coupled-wire construction can also be generalized to the systems with arbitrary cyrstalline symmetry $SG$ and internal symmetry $G_0$ in arbitrary dimensions, and especially in 2D Moir\'e superlattice, with more complicated inter-wire entanglement patterns, and their quantum phase transitions should also be controlled by inter-wire entanglement patterns of quantum wires. Furthermore, with explicit corner modes, the 2D HOTSC may directly justified by tunneling spectroscopic measurements on the edge.

\textit{Acknowledgements} --   The author thanks Meng Cheng, Shang-Qiang Ning, Liujun Zou and Rui-Xing Zhang for stimulating discussions. This work is supported by Direct
Grant No. 4053462 from The Chinese University of Hong Kong and funding from Hong Kong’s Research Grants Council (GRF No. 14306420, ANR/RGC Joint Research Scheme No. A-CUHK402/18).

\providecommand{\noopsort}[1]{}\providecommand{\singleletter}[1]{#1}%
%


\begin{thebibliography}{92}%
\makeatletter
\providecommand \@ifxundefined [1]{%
 \@ifx{#1\undefined}
}%
\providecommand \@ifnum [1]{%
 \ifnum #1\expandafter \@firstoftwo
 \else \expandafter \@secondoftwo
 \fi
}%
\providecommand \@ifx [1]{%
 \ifx #1\expandafter \@firstoftwo
 \else \expandafter \@secondoftwo
 \fi
}%
\providecommand \natexlab [1]{#1}%
\providecommand \enquote  [1]{``#1''}%
\providecommand \bibnamefont  [1]{#1}%
\providecommand \bibfnamefont [1]{#1}%
\providecommand \citenamefont [1]{#1}%
\providecommand \href@noop [0]{\@secondoftwo}%
\providecommand \href [0]{\begingroup \@sanitize@url \@href}%
\providecommand \@href[1]{\@@startlink{#1}\@@href}%
\providecommand \@@href[1]{\endgroup#1\@@endlink}%
\providecommand \@sanitize@url [0]{\catcode `\\12\catcode `\$12\catcode
  `\&12\catcode `\#12\catcode `\^12\catcode `\_12\catcode `\%12\relax}%
\providecommand \@@startlink[1]{}%
\providecommand \@@endlink[0]{}%
\providecommand \url  [0]{\begingroup\@sanitize@url \@url }%
\providecommand \@url [1]{\endgroup\@href {#1}{\urlprefix }}%
\providecommand \urlprefix  [0]{URL }%
\providecommand \Eprint [0]{\href }%
\providecommand \doibase [0]{http://dx.doi.org/}%
\providecommand \selectlanguage [0]{\@gobble}%
\providecommand \bibinfo  [0]{\@secondoftwo}%
\providecommand \bibfield  [0]{\@secondoftwo}%
\providecommand \translation [1]{[#1]}%
\providecommand \BibitemOpen [0]{}%
\providecommand \bibitemStop [0]{}%
\providecommand \bibitemNoStop [0]{.\EOS\space}%
\providecommand \EOS [0]{\spacefactor3000\relax}%
\providecommand \BibitemShut  [1]{\csname bibitem#1\endcsname}%
\let\auto@bib@innerbib\@empty
\bibitem [{\citenamefont {Tsui}\ \emph {et~al.}(1982)\citenamefont {Tsui},
  \citenamefont {Stormer},\ and\ \citenamefont {Gossard}}]{FQHE}%
  \BibitemOpen
  \bibfield  {author} {\bibinfo {author} {\bibfnamefont {D.~C.}\ \bibnamefont
  {Tsui}}, \bibinfo {author} {\bibfnamefont {H.~L.}\ \bibnamefont {Stormer}}, \
  and\ \bibinfo {author} {\bibfnamefont {A.~C.}\ \bibnamefont {Gossard}},\
  }\bibfield  {title} {\enquote {\bibinfo {title} {Two-dimensional
  magnetotransport in the extreme quantum limit},}\ }\href {\doibase
  10.1103/PhysRevLett.48.1559} {\bibfield  {journal} {\bibinfo  {journal}
  {Phys. Rev. Lett.}\ }\textbf {\bibinfo {volume} {48}},\ \bibinfo {pages}
  {1559} (\bibinfo {year} {1982})}\BibitemShut {NoStop}%
\bibitem [{\citenamefont {Laughlin}(1983)}]{Laughlin}%
  \BibitemOpen
  \bibfield  {author} {\bibinfo {author} {\bibfnamefont {R.~B.}\ \bibnamefont
  {Laughlin}},\ }\bibfield  {title} {\enquote {\bibinfo {title} {Anomalous
  quantum hall effect: An incompressible quantum fluid with fractionally
  charged excitations},}\ }\href@noop {} {\bibfield  {journal} {\bibinfo
  {journal} {Phys. Rev. Lett.}\ }\textbf {\bibinfo {volume} {50}},\ \bibinfo
  {pages} {1395} (\bibinfo {year} {1983})}\BibitemShut {NoStop}%
\bibitem [{\citenamefont {Chen}\ \emph {et~al.}(2010)\citenamefont {Chen},
  \citenamefont {Gu},\ and\ \citenamefont {Wen}}]{entanglement}%
  \BibitemOpen
  \bibfield  {author} {\bibinfo {author} {\bibfnamefont {X.}~\bibnamefont
  {Chen}}, \bibinfo {author} {\bibfnamefont {Z.-C.}\ \bibnamefont {Gu}}, \ and\
  \bibinfo {author} {\bibfnamefont {X.-G.}\ \bibnamefont {Wen}},\ }\bibfield
  {title} {\enquote {\bibinfo {title} {Local unitary transformation, long-range
  quantum entanglement, wave function renormalization, and topological
  order},}\ }\href
  {https://journals.aps.org/prb/abstract/10.1103/PhysRevB.82.155138} {\bibfield
   {journal} {\bibinfo  {journal} {Phys. Rev. B}\ }\textbf {\bibinfo {volume}
  {82}},\ \bibinfo {pages} {155138} (\bibinfo {year} {2010})}\BibitemShut
  {NoStop}%
\bibitem [{\citenamefont {Gu}\ and\ \citenamefont {Wen}(2009)}]{ZCGu2009}%
  \BibitemOpen
  \bibfield  {author} {\bibinfo {author} {\bibfnamefont {Z.-C.}\ \bibnamefont
  {Gu}}\ and\ \bibinfo {author} {\bibfnamefont {X.-G.}\ \bibnamefont {Wen}},\
  }\bibfield  {title} {\enquote {\bibinfo {title}
  {Tensor-entanglement-filtering renormalization approach and
  symmetry-protected topological order},}\ }\href
  {https://journals.aps.org/prb/abstract/10.1103/PhysRevB.80.155131} {\bibfield
   {journal} {\bibinfo  {journal} {Phys. Rev. B}\ }\textbf {\bibinfo {volume}
  {80}},\ \bibinfo {pages} {155131} (\bibinfo {year} {2009})}\BibitemShut
  {NoStop}%
\bibitem [{\citenamefont {Chen}\ \emph {et~al.}(2011)\citenamefont {Chen},
  \citenamefont {Gu},\ and\ \citenamefont {Wen}}]{chen11a}%
  \BibitemOpen
  \bibfield  {author} {\bibinfo {author} {\bibfnamefont {X.}~\bibnamefont
  {Chen}}, \bibinfo {author} {\bibfnamefont {Z.-C.}\ \bibnamefont {Gu}}, \ and\
  \bibinfo {author} {\bibfnamefont {X.-G.}\ \bibnamefont {Wen}},\ }\bibfield
  {title} {\enquote {\bibinfo {title} {Classification of gapped symmetric
  phases in one-dimensional spin systems},}\ }\href {\doibase
  10.1103/PhysRevB.83.035107} {\bibfield  {journal} {\bibinfo  {journal} {Phys.
  Rev. B}\ }\textbf {\bibinfo {volume} {83}},\ \bibinfo {pages} {035107}
  (\bibinfo {year} {2011})}\BibitemShut {NoStop}%
\bibitem [{\citenamefont {Chen}\ \emph {et~al.}(2012)\citenamefont {Chen},
  \citenamefont {Gu}, \citenamefont {Liu},\ and\ \citenamefont
  {Wen}}]{XieChenScience}%
  \BibitemOpen
  \bibfield  {author} {\bibinfo {author} {\bibfnamefont {X.}~\bibnamefont
  {Chen}}, \bibinfo {author} {\bibfnamefont {Z.-C.}\ \bibnamefont {Gu}},
  \bibinfo {author} {\bibfnamefont {Z.-X.}\ \bibnamefont {Liu}}, \ and\
  \bibinfo {author} {\bibfnamefont {X.-G.}\ \bibnamefont {Wen}},\ }\bibfield
  {title} {\enquote {\bibinfo {title} {Symmetry-protected topological orders in
  interacting bosonic systems},}\ }\href
  {https://science.sciencemag.org/content/338/6114/1604} {\bibfield  {journal}
  {\bibinfo  {journal} {Science}\ }\textbf {\bibinfo {volume} {338}},\ \bibinfo
  {pages} {1604--1606} (\bibinfo {year} {2012})}\BibitemShut {NoStop}%
\bibitem [{\citenamefont {Chen}\ \emph {et~al.}(2013)\citenamefont {Chen},
  \citenamefont {Gu}, \citenamefont {Liu},\ and\ \citenamefont
  {Wen}}]{cohomology}%
  \BibitemOpen
  \bibfield  {author} {\bibinfo {author} {\bibfnamefont {X.}~\bibnamefont
  {Chen}}, \bibinfo {author} {\bibfnamefont {Z.-C.}\ \bibnamefont {Gu}},
  \bibinfo {author} {\bibfnamefont {Z.-X.}\ \bibnamefont {Liu}}, \ and\
  \bibinfo {author} {\bibfnamefont {X.-G.}\ \bibnamefont {Wen}},\ }\bibfield
  {title} {\enquote {\bibinfo {title} {Symmetry protected topological orders
  and the group cohomology of their symmetry group},}\ }\href
  {https://journals.aps.org/prb/abstract/10.1103/PhysRevB.87.155114} {\bibfield
   {journal} {\bibinfo  {journal} {Phys. Rev. B}\ }\textbf {\bibinfo {volume}
  {87}},\ \bibinfo {pages} {155114} (\bibinfo {year} {2013})}\BibitemShut
  {NoStop}%
\bibitem [{\citenamefont {Senthil}(2015)}]{Senthil_2015}%
  \BibitemOpen
  \bibfield  {author} {\bibinfo {author} {\bibfnamefont {T.}~\bibnamefont
  {Senthil}},\ }\bibfield  {title} {\enquote {\bibinfo {title}
  {Symmetry-protected topological phases of quantum matter},}\ }\href {\doibase
  10.1146/annurev-conmatphys-031214-014740} {\bibfield  {journal} {\bibinfo
  {journal} {Annu. Rev. Condens. Matter Phys.}\ }\textbf {\bibinfo {volume}
  {6}},\ \bibinfo {pages} {299--324} (\bibinfo {year} {2015})}\BibitemShut
  {NoStop}%
\bibitem [{\citenamefont {Plamadeala}\ \emph {et~al.}(2012)\citenamefont
  {Plamadeala}, \citenamefont {Mulligan},\ and\ \citenamefont {Nayak}}]{E8}%
  \BibitemOpen
  \bibfield  {author} {\bibinfo {author} {\bibfnamefont {E.}~\bibnamefont
  {Plamadeala}}, \bibinfo {author} {\bibfnamefont {M.}~\bibnamefont
  {Mulligan}}, \ and\ \bibinfo {author} {\bibfnamefont {C.}~\bibnamefont
  {Nayak}},\ }\bibfield  {title} {\enquote {\bibinfo {title} {Short-range
  entangled bosonic states with chiral edge modes and t duality of heterotic
  strings},}\ }\href
  {https://journals.aps.org/prb/abstract/10.1103/PhysRevB.88.045131} {\bibfield
   {journal} {\bibinfo  {journal} {Phys. Rev. B}\ }\textbf {\bibinfo {volume}
  {88}},\ \bibinfo {pages} {045131} (\bibinfo {year} {2012})}\BibitemShut
  {NoStop}%
\bibitem [{\citenamefont {Lu}\ and\ \citenamefont {Vishwanath}(2012)}]{Lu12}%
  \BibitemOpen
  \bibfield  {author} {\bibinfo {author} {\bibfnamefont {Yuan-Ming}\
  \bibnamefont {Lu}}\ and\ \bibinfo {author} {\bibfnamefont {Ashvin}\
  \bibnamefont {Vishwanath}},\ }\bibfield  {title} {\enquote {\bibinfo {title}
  {Theory and classification of interacting integer topological phases in two
  dimensions: A chern-simons approach},}\ }\href {\doibase
  10.1103/PhysRevB.86.125119} {\bibfield  {journal} {\bibinfo  {journal} {Phys.
  Rev. B}\ }\textbf {\bibinfo {volume} {86}},\ \bibinfo {pages} {125119}
  (\bibinfo {year} {2012})}\BibitemShut {NoStop}%
\bibitem [{\citenamefont {Freed}()}]{invertible2}%
  \BibitemOpen
  \bibfield  {author} {\bibinfo {author} {\bibfnamefont {D.~S.}\ \bibnamefont
  {Freed}},\ }\bibfield  {title} {\enquote {\bibinfo {title} {Short-range
  entanglement and invertible field theories},}\ }\href@noop {} {\ }\Eprint
  {http://arxiv.org/abs/1406.7278} {arXiv:1406.7278 [cond-mat.str-el]}
  \BibitemShut {NoStop}%
\bibitem [{\citenamefont {Freed}\ and\ \citenamefont
  {Hopkins}(2016)}]{invertible3}%
  \BibitemOpen
  \bibfield  {author} {\bibinfo {author} {\bibfnamefont {Daniel~S.}\
  \bibnamefont {Freed}}\ and\ \bibinfo {author} {\bibfnamefont {Michael~J.}\
  \bibnamefont {Hopkins}},\ }\bibfield  {title} {\enquote {\bibinfo {title}
  {{Reflection positivity and invertible topological phases}},}\ }\href@noop {}
  {\bibfield  {journal} {\bibinfo  {journal} {arXiv e-prints}\ } (\bibinfo
  {year} {2016})},\ \Eprint {http://arxiv.org/abs/1604.06527}
  {arXiv:1604.06527} \BibitemShut {NoStop}%
\bibitem [{\citenamefont {Gu}\ and\ \citenamefont {Wen}(2014)}]{special}%
  \BibitemOpen
  \bibfield  {author} {\bibinfo {author} {\bibfnamefont {Z.-C.}\ \bibnamefont
  {Gu}}\ and\ \bibinfo {author} {\bibfnamefont {X.-G.}\ \bibnamefont {Wen}},\
  }\bibfield  {title} {\enquote {\bibinfo {title} {Symmetry-protected
  topological orders for interacting fermions: Fermionic topological nonlinear
  $\sigma$ models and a special group supercohomology theory},}\ }\href
  {\doibase 10.1103/PhysRevB.90.115141} {\bibfield  {journal} {\bibinfo
  {journal} {Phys. Rev. B}\ }\textbf {\bibinfo {volume} {90}},\ \bibinfo
  {pages} {115141} (\bibinfo {year} {2014})}\BibitemShut {NoStop}%
\bibitem [{\citenamefont {Wang}\ and\ \citenamefont {Gu}(2018)}]{general1}%
  \BibitemOpen
  \bibfield  {author} {\bibinfo {author} {\bibfnamefont {Q.-R.}\ \bibnamefont
  {Wang}}\ and\ \bibinfo {author} {\bibfnamefont {Z.-C.}\ \bibnamefont {Gu}},\
  }\bibfield  {title} {\enquote {\bibinfo {title} {Towards a complete
  classification of symmetry-protected topological phases for interacting
  fermions in three dimensions and a general group supercohomology theory},}\
  }\href {https://journals.aps.org/prx/abstract/10.1103/PhysRevX.8.011055}
  {\bibfield  {journal} {\bibinfo  {journal} {Phys. Rev. X}\ }\textbf {\bibinfo
  {volume} {8}},\ \bibinfo {pages} {011055} (\bibinfo {year}
  {2018})}\BibitemShut {NoStop}%
\bibitem [{\citenamefont {Wang}\ and\ \citenamefont {Gu}(2020)}]{general2}%
  \BibitemOpen
  \bibfield  {author} {\bibinfo {author} {\bibfnamefont {Q.-R.}\ \bibnamefont
  {Wang}}\ and\ \bibinfo {author} {\bibfnamefont {Z.-C.}\ \bibnamefont {Gu}},\
  }\bibfield  {title} {\enquote {\bibinfo {title} {Construction and
  classification of symmetry-protected topological phases in interacting
  fermion systems},}\ }\href {\doibase 10.1103/PhysRevX.10.031055} {\bibfield
  {journal} {\bibinfo  {journal} {Phys. Rev. X}\ }\textbf {\bibinfo {volume}
  {10}},\ \bibinfo {pages} {031055} (\bibinfo {year} {2020})},\ \Eprint
  {http://arxiv.org/abs/1811.00536} {arXiv:1811.00536 [cond-mat.str-el]}
  \BibitemShut {NoStop}%
\bibitem [{\citenamefont {Kapustin}()}]{Kapustin2014}%
  \BibitemOpen
  \bibfield  {author} {\bibinfo {author} {\bibfnamefont {A.}~\bibnamefont
  {Kapustin}},\ }\bibfield  {title} {\enquote {\bibinfo {title} {Symmetry
  protected topological phases, anomalies, and cobordisms: Beyond group
  cohomology},}\ }\href@noop {} {\ }\Eprint {http://arxiv.org/abs/1403.1467}
  {arXiv:1403.1467 [cond-mat.str-el]} \BibitemShut {NoStop}%
\bibitem [{\citenamefont {Kapustin}\ \emph {et~al.}(2015)\citenamefont
  {Kapustin}, \citenamefont {Thorngren}, \citenamefont {Turzillo},\ and\
  \citenamefont {Wang}}]{Kapustin2015}%
  \BibitemOpen
  \bibfield  {author} {\bibinfo {author} {\bibfnamefont {Anton}\ \bibnamefont
  {Kapustin}}, \bibinfo {author} {\bibfnamefont {Ryan}\ \bibnamefont
  {Thorngren}}, \bibinfo {author} {\bibfnamefont {Alex}\ \bibnamefont
  {Turzillo}}, \ and\ \bibinfo {author} {\bibfnamefont {Zitao}\ \bibnamefont
  {Wang}},\ }\bibfield  {title} {\enquote {\bibinfo {title} {Fermionic symmetry
  protected topological phases and cobordisms},}\ }\href {\doibase
  https://link.springer.com/article/10.1007/JHEP12(2015)052} {\bibfield
  {journal} {\bibinfo  {journal} {JHEP}\ }\textbf {\bibinfo {volume} {1512}},\
  \bibinfo {pages} {052} (\bibinfo {year} {2015})}\BibitemShut {NoStop}%
\bibitem [{\citenamefont {Kapustin}\ and\ \citenamefont
  {Thorngren}(2017)}]{Kapustin2017}%
  \BibitemOpen
  \bibfield  {author} {\bibinfo {author} {\bibfnamefont {Anton}\ \bibnamefont
  {Kapustin}}\ and\ \bibinfo {author} {\bibfnamefont {Ryan}\ \bibnamefont
  {Thorngren}},\ }\bibfield  {title} {\enquote {\bibinfo {title} {Fermionic spt
  phases in higher dimensions and bosonization},}\ }\href {\doibase
  10.1007/JHEP10(2017)080} {\bibfield  {journal} {\bibinfo  {journal} {Journal
  of High Energy Physics}\ }\textbf {\bibinfo {volume} {2017}},\ \bibinfo
  {pages} {80} (\bibinfo {year} {2017})}\BibitemShut {NoStop}%
\bibitem [{\citenamefont {Gu}\ and\ \citenamefont {Levin}(2014)}]{Gu-Levin}%
  \BibitemOpen
  \bibfield  {author} {\bibinfo {author} {\bibfnamefont {Z.-C.}\ \bibnamefont
  {Gu}}\ and\ \bibinfo {author} {\bibfnamefont {M.}~\bibnamefont {Levin}},\
  }\bibfield  {title} {\enquote {\bibinfo {title} {Effect of interactions on
  two-dimensional fermionic symmetry-protected topological phases with $z_2$
  symmetry},}\ }\href
  {https://journals.aps.org/prb/abstract/10.1103/PhysRevB.89.201113} {\bibfield
   {journal} {\bibinfo  {journal} {Phys. Rev. B}\ }\textbf {\bibinfo {volume}
  {89}},\ \bibinfo {pages} {201113(R)} (\bibinfo {year} {2014})}\BibitemShut
  {NoStop}%
\bibitem [{\citenamefont {Cheng}\ and\ \citenamefont {Gu}(2014)}]{gauging1}%
  \BibitemOpen
  \bibfield  {author} {\bibinfo {author} {\bibfnamefont {M.}~\bibnamefont
  {Cheng}}\ and\ \bibinfo {author} {\bibfnamefont {Z.-C.}\ \bibnamefont {Gu}},\
  }\bibfield  {title} {\enquote {\bibinfo {title} {Topological response theory
  of abelian symmetry-protected topological phases in two dimensions},}\ }\href
  {https://journals.aps.org/prl/abstract/10.1103/PhysRevLett.112.141602}
  {\bibfield  {journal} {\bibinfo  {journal} {Phys. Rev. Lett.}\ }\textbf
  {\bibinfo {volume} {112}},\ \bibinfo {pages} {141602} (\bibinfo {year}
  {2014})}\BibitemShut {NoStop}%
\bibitem [{\citenamefont {Barkeshli}\ \emph {et~al.}(2019)\citenamefont
  {Barkeshli}, \citenamefont {Bonderson}, \citenamefont {Cheng},\ and\
  \citenamefont {Wang}}]{gauging3}%
  \BibitemOpen
  \bibfield  {author} {\bibinfo {author} {\bibfnamefont {M.}~\bibnamefont
  {Barkeshli}}, \bibinfo {author} {\bibfnamefont {P.}~\bibnamefont
  {Bonderson}}, \bibinfo {author} {\bibfnamefont {M.}~\bibnamefont {Cheng}}, \
  and\ \bibinfo {author} {\bibfnamefont {Z.}~\bibnamefont {Wang}},\ }\bibfield
  {title} {\enquote {\bibinfo {title} {Symmetry fractionalization, defects, and
  gauging of topological phases},}\ }\href {\doibase
  10.1103/PhysRevB.100.115147} {\bibfield  {journal} {\bibinfo  {journal}
  {Phys. Rev. B}\ }\textbf {\bibinfo {volume} {100}},\ \bibinfo {pages}
  {115147} (\bibinfo {year} {2019})},\ \Eprint {http://arxiv.org/abs/1410.4540}
  {arXiv:1410.4540 [cond-mat.str-el]} \BibitemShut {NoStop}%
\bibitem [{\citenamefont {Tantivasadakarn}(2017)}]{dimensionalreduction}%
  \BibitemOpen
  \bibfield  {author} {\bibinfo {author} {\bibfnamefont {N.}~\bibnamefont
  {Tantivasadakarn}},\ }\bibfield  {title} {\enquote {\bibinfo {title}
  {Dimensional reduction and topological invariants of symmetry-protected
  topological phases},}\ }\href
  {https://journals.aps.org/prb/abstract/10.1103/PhysRevB.96.195101} {\bibfield
   {journal} {\bibinfo  {journal} {Phys. Rev. B}\ }\textbf {\bibinfo {volume}
  {96}},\ \bibinfo {pages} {195101} (\bibinfo {year} {2017})}\BibitemShut
  {NoStop}%
\bibitem [{\citenamefont {Wang}\ \emph {et~al.}(2017)\citenamefont {Wang},
  \citenamefont {Lin},\ and\ \citenamefont {Gu}}]{gauging2}%
  \BibitemOpen
  \bibfield  {author} {\bibinfo {author} {\bibfnamefont {C.}~\bibnamefont
  {Wang}}, \bibinfo {author} {\bibfnamefont {C.-H.}\ \bibnamefont {Lin}}, \
  and\ \bibinfo {author} {\bibfnamefont {Z.-C.}\ \bibnamefont {Gu}},\
  }\bibfield  {title} {\enquote {\bibinfo {title} {Interacting fermionic
  symmetry-protected topological phases in two dimensions},}\ }\href
  {https://journals.aps.org/prb/abstract/10.1103/PhysRevB.95.195147} {\bibfield
   {journal} {\bibinfo  {journal} {Phys. Rev. B}\ }\textbf {\bibinfo {volume}
  {95}},\ \bibinfo {pages} {195147} (\bibinfo {year} {2017})}\BibitemShut
  {NoStop}%
\bibitem [{\citenamefont {Cheng}\ \emph
  {et~al.}(2018{\natexlab{a}})\citenamefont {Cheng}, \citenamefont {Bi},
  \citenamefont {You},\ and\ \citenamefont {Gu}}]{2DFSPT}%
  \BibitemOpen
  \bibfield  {author} {\bibinfo {author} {\bibfnamefont {M.}~\bibnamefont
  {Cheng}}, \bibinfo {author} {\bibfnamefont {Z.}~\bibnamefont {Bi}}, \bibinfo
  {author} {\bibfnamefont {Y.-Z.}\ \bibnamefont {You}}, \ and\ \bibinfo
  {author} {\bibfnamefont {Z.-C.}\ \bibnamefont {Gu}},\ }\bibfield  {title}
  {\enquote {\bibinfo {title} {Classification of symmetry-protected phases for
  interacting fermions in two dimensions},}\ }\href
  {https://journals.aps.org/prb/abstract/10.1103/PhysRevB.97.205109} {\bibfield
   {journal} {\bibinfo  {journal} {Phys. Rev. B}\ }\textbf {\bibinfo {volume}
  {97}},\ \bibinfo {pages} {205109} (\bibinfo {year}
  {2018}{\natexlab{a}})}\BibitemShut {NoStop}%
\bibitem [{\citenamefont {Cheng}\ \emph
  {et~al.}(2018{\natexlab{b}})\citenamefont {Cheng}, \citenamefont
  {Tantivasadakarn},\ and\ \citenamefont {Wang}}]{braiding}%
  \BibitemOpen
  \bibfield  {author} {\bibinfo {author} {\bibfnamefont {M.}~\bibnamefont
  {Cheng}}, \bibinfo {author} {\bibfnamefont {N.}~\bibnamefont
  {Tantivasadakarn}}, \ and\ \bibinfo {author} {\bibfnamefont {C.}~\bibnamefont
  {Wang}},\ }\bibfield  {title} {\enquote {\bibinfo {title} {Loop braiding
  statistics and interacting fermionic symmetry-protected topological phases in
  three dimensions},}\ }\href
  {https://journals.aps.org/prx/abstract/10.1103/PhysRevX.8.011054} {\bibfield
  {journal} {\bibinfo  {journal} {Phys. Rev. X}\ }\textbf {\bibinfo {volume}
  {8}},\ \bibinfo {pages} {011054} (\bibinfo {year}
  {2018}{\natexlab{b}})}\BibitemShut {NoStop}%
\bibitem [{\citenamefont {Ning}\ \emph {et~al.}(2021)\citenamefont {Ning},
  \citenamefont {Wang}, \citenamefont {Wang},\ and\ \citenamefont
  {Gu}}]{Ning21a}%
  \BibitemOpen
  \bibfield  {author} {\bibinfo {author} {\bibfnamefont {Shang-Qiang}\
  \bibnamefont {Ning}}, \bibinfo {author} {\bibfnamefont {Chenjie}\
  \bibnamefont {Wang}}, \bibinfo {author} {\bibfnamefont {Qing-Rui}\
  \bibnamefont {Wang}}, \ and\ \bibinfo {author} {\bibfnamefont {Zheng-Cheng}\
  \bibnamefont {Gu}},\ }\bibfield  {title} {\enquote {\bibinfo {title} {Edge
  theories of two-dimensional fermionic symmetry protected topological phases
  protected by unitary abelian symmetries},}\ }\href {\doibase
  10.1103/PhysRevB.104.075151} {\bibfield  {journal} {\bibinfo  {journal}
  {Phys. Rev. B}\ }\textbf {\bibinfo {volume} {104}},\ \bibinfo {pages}
  {075151} (\bibinfo {year} {2021})}\BibitemShut {NoStop}%
\bibitem [{\citenamefont {Hasan}\ and\ \citenamefont {Kane}(2010)}]{KaneRMP}%
  \BibitemOpen
  \bibfield  {author} {\bibinfo {author} {\bibfnamefont {M.~Z.}\ \bibnamefont
  {Hasan}}\ and\ \bibinfo {author} {\bibfnamefont {C.~L.}\ \bibnamefont
  {Kane}},\ }\bibfield  {title} {\enquote {\bibinfo {title} {Colloquium:
  Topological insulators},}\ }\href
  {https://journals.aps.org/rmp/abstract/10.1103/RevModPhys.82.3045} {\bibfield
   {journal} {\bibinfo  {journal} {Rev. Mod. Phys.}\ }\textbf {\bibinfo
  {volume} {82}},\ \bibinfo {pages} {3045--3067} (\bibinfo {year}
  {2010})}\BibitemShut {NoStop}%
\bibitem [{\citenamefont {Qi}\ and\ \citenamefont {Zhang}(2011)}]{ZhangRMP}%
  \BibitemOpen
  \bibfield  {author} {\bibinfo {author} {\bibfnamefont {X.-L.}\ \bibnamefont
  {Qi}}\ and\ \bibinfo {author} {\bibfnamefont {S.-C.}\ \bibnamefont {Zhang}},\
  }\bibfield  {title} {\enquote {\bibinfo {title} {Topological insulators and
  superconductors},}\ }\href
  {https://journals.aps.org/rmp/abstract/10.1103/RevModPhys.83.1057} {\bibfield
   {journal} {\bibinfo  {journal} {Rev. Mod. Phys.}\ }\textbf {\bibinfo
  {volume} {83}},\ \bibinfo {pages} {1057--1110} (\bibinfo {year}
  {2011})}\BibitemShut {NoStop}%
\bibitem [{\citenamefont {Fu}(2011)}]{TCI}%
  \BibitemOpen
  \bibfield  {author} {\bibinfo {author} {\bibfnamefont {L.}~\bibnamefont
  {Fu}},\ }\bibfield  {title} {\enquote {\bibinfo {title} {Topological
  crystalline insulators},}\ }\href
  {https://journals.aps.org/prl/abstract/10.1103/PhysRevLett.106.106802}
  {\bibfield  {journal} {\bibinfo  {journal} {Phys. Rev. Lett.}\ }\textbf
  {\bibinfo {volume} {106}},\ \bibinfo {pages} {106802} (\bibinfo {year}
  {2011})}\BibitemShut {NoStop}%
\bibitem [{\citenamefont {Hsieh}\ \emph {et~al.}(2012)\citenamefont {Hsieh},
  \citenamefont {Lin}, \citenamefont {Liu}, \citenamefont {Duan}, \citenamefont
  {Bansil},\ and\ \citenamefont {Fu}}]{Fu2012}%
  \BibitemOpen
  \bibfield  {author} {\bibinfo {author} {\bibfnamefont {T.~H.}\ \bibnamefont
  {Hsieh}}, \bibinfo {author} {\bibfnamefont {H.}~\bibnamefont {Lin}}, \bibinfo
  {author} {\bibfnamefont {J.}~\bibnamefont {Liu}}, \bibinfo {author}
  {\bibfnamefont {W.}~\bibnamefont {Duan}}, \bibinfo {author} {\bibfnamefont
  {A.}~\bibnamefont {Bansil}}, \ and\ \bibinfo {author} {\bibfnamefont
  {L.}~\bibnamefont {Fu}},\ }\bibfield  {title} {\enquote {\bibinfo {title}
  {Topological crystalline insulators in the snte material class},}\ }\href
  {\doibase 10.1038/ncomms1969} {\bibfield  {journal} {\bibinfo  {journal}
  {Nat. Commun.}\ }\textbf {\bibinfo {volume} {3}},\ \bibinfo {pages} {982}
  (\bibinfo {year} {2012})}\BibitemShut {NoStop}%
\bibitem [{\citenamefont {Isobe}\ and\ \citenamefont {Fu}(2015)}]{ITCI}%
  \BibitemOpen
  \bibfield  {author} {\bibinfo {author} {\bibfnamefont {H.}~\bibnamefont
  {Isobe}}\ and\ \bibinfo {author} {\bibfnamefont {L.}~\bibnamefont {Fu}},\
  }\bibfield  {title} {\enquote {\bibinfo {title} {Theory of interacting
  topological crystalline insulators},}\ }\href
  {https://journals.aps.org/prb/abstract/10.1103/PhysRevB.92.081304} {\bibfield
   {journal} {\bibinfo  {journal} {Phys. Rev. B}\ }\textbf {\bibinfo {volume}
  {92}},\ \bibinfo {pages} {081304(R)} (\bibinfo {year} {2015})}\BibitemShut
  {NoStop}%
\bibitem [{\citenamefont {Song}\ \emph {et~al.}(2017)\citenamefont {Song},
  \citenamefont {Huang}, \citenamefont {Fu},\ and\ \citenamefont
  {Hermele}}]{reduction}%
  \BibitemOpen
  \bibfield  {author} {\bibinfo {author} {\bibfnamefont {H.}~\bibnamefont
  {Song}}, \bibinfo {author} {\bibfnamefont {S.-J.}\ \bibnamefont {Huang}},
  \bibinfo {author} {\bibfnamefont {L.}~\bibnamefont {Fu}}, \ and\ \bibinfo
  {author} {\bibfnamefont {M.}~\bibnamefont {Hermele}},\ }\bibfield  {title}
  {\enquote {\bibinfo {title} {Topological phases protected by point group
  symmetry},}\ }\href
  {https://journals.aps.org/prx/abstract/10.1103/PhysRevX.7.011020} {\bibfield
  {journal} {\bibinfo  {journal} {Phys. Rev. X}\ }\textbf {\bibinfo {volume}
  {7}},\ \bibinfo {pages} {011020} (\bibinfo {year} {2017})}\BibitemShut
  {NoStop}%
\bibitem [{\citenamefont {Huang}\ \emph {et~al.}(2017)\citenamefont {Huang},
  \citenamefont {Song}, \citenamefont {Huang},\ and\ \citenamefont
  {Hermele}}]{building}%
  \BibitemOpen
  \bibfield  {author} {\bibinfo {author} {\bibfnamefont {S.-J.}\ \bibnamefont
  {Huang}}, \bibinfo {author} {\bibfnamefont {H.}~\bibnamefont {Song}},
  \bibinfo {author} {\bibfnamefont {Y.-P.}\ \bibnamefont {Huang}}, \ and\
  \bibinfo {author} {\bibfnamefont {M.}~\bibnamefont {Hermele}},\ }\bibfield
  {title} {\enquote {\bibinfo {title} {Building crystalline topological phases
  from lower-dimensional states},}\ }\href
  {https://journals.aps.org/prb/abstract/10.1103/PhysRevB.96.205106} {\bibfield
   {journal} {\bibinfo  {journal} {Phys. Rev. B}\ }\textbf {\bibinfo {volume}
  {96}},\ \bibinfo {pages} {205106} (\bibinfo {year} {2017})}\BibitemShut
  {NoStop}%
\bibitem [{\citenamefont {Thorngren}\ and\ \citenamefont
  {Else}(2018)}]{correspondence}%
  \BibitemOpen
  \bibfield  {author} {\bibinfo {author} {\bibfnamefont {Ryan}\ \bibnamefont
  {Thorngren}}\ and\ \bibinfo {author} {\bibfnamefont {Dominic~V.}\
  \bibnamefont {Else}},\ }\bibfield  {title} {\enquote {\bibinfo {title}
  {Gauging spatial symmetries and the classification of topological crystalline
  phases},}\ }\href
  {https://journals.aps.org/prx/abstract/10.1103/PhysRevX.8.011040} {\bibfield
  {journal} {\bibinfo  {journal} {Phys. Rev. X}\ }\textbf {\bibinfo {volume}
  {8}},\ \bibinfo {pages} {011040} (\bibinfo {year} {2018})}\BibitemShut
  {NoStop}%
\bibitem [{\citenamefont {Zou}(2018)}]{SET}%
  \BibitemOpen
  \bibfield  {author} {\bibinfo {author} {\bibfnamefont {L.}~\bibnamefont
  {Zou}},\ }\bibfield  {title} {\enquote {\bibinfo {title} {Bulk
  characterization of topological crystalline insulators: Stability under
  interactions and relations to symmetry enriched $u(1)$ quantum spin
  liquids},}\ }\href
  {https://journals.aps.org/prb/abstract/10.1103/PhysRevB.97.045130} {\bibfield
   {journal} {\bibinfo  {journal} {Phys. Rev. B}\ }\textbf {\bibinfo {volume}
  {97}},\ \bibinfo {pages} {045130} (\bibinfo {year} {2018})}\BibitemShut
  {NoStop}%
\bibitem [{\citenamefont {Po}\ \emph {et~al.}(2017)\citenamefont {Po},
  \citenamefont {Vishwanath},\ and\ \citenamefont {Watanabe}}]{230}%
  \BibitemOpen
  \bibfield  {author} {\bibinfo {author} {\bibfnamefont {H.~C.}\ \bibnamefont
  {Po}}, \bibinfo {author} {\bibfnamefont {A.}~\bibnamefont {Vishwanath}}, \
  and\ \bibinfo {author} {\bibfnamefont {H.}~\bibnamefont {Watanabe}},\
  }\bibfield  {title} {\enquote {\bibinfo {title} {Symmetry-based indicators of
  band topology in the 230 space groups},}\ }\href
  {https://www.nature.com/articles/s41467-017-00133-2} {\bibfield  {journal}
  {\bibinfo  {journal} {Nature Communications}\ }\textbf {\bibinfo {volume}
  {8}},\ \bibinfo {pages} {50} (\bibinfo {year} {2017})}\BibitemShut {NoStop}%
\bibitem [{\citenamefont {Song}\ \emph
  {et~al.}(2020{\natexlab{a}})\citenamefont {Song}, \citenamefont {Xiong},\
  and\ \citenamefont {Huang}}]{BCSPT}%
  \BibitemOpen
  \bibfield  {author} {\bibinfo {author} {\bibfnamefont {H.}~\bibnamefont
  {Song}}, \bibinfo {author} {\bibfnamefont {C.~Z.}\ \bibnamefont {Xiong}}, \
  and\ \bibinfo {author} {\bibfnamefont {S.-J.}\ \bibnamefont {Huang}},\
  }\bibfield  {title} {\enquote {\bibinfo {title} {Bosonic crystalline symmetry
  protected topological phases beyond the group cohomology proposal},}\ }\href
  {\doibase 10.1103/PhysRevB.101.165129} {\bibfield  {journal} {\bibinfo
  {journal} {Phys. Rev. B}\ }\textbf {\bibinfo {volume} {101}},\ \bibinfo
  {pages} {165129} (\bibinfo {year} {2020}{\natexlab{a}})},\ \Eprint
  {http://arxiv.org/abs/1811.06558} {arXiv:1811.06558 [cond-mat.str-el]}
  \BibitemShut {NoStop}%
\bibitem [{\citenamefont {Jiang}\ and\ \citenamefont {Ran}(2017)}]{Jiang2017}%
  \BibitemOpen
  \bibfield  {author} {\bibinfo {author} {\bibfnamefont {S.}~\bibnamefont
  {Jiang}}\ and\ \bibinfo {author} {\bibfnamefont {Y.}~\bibnamefont {Ran}},\
  }\bibfield  {title} {\enquote {\bibinfo {title} {Anyon condensation and a
  generic tensor-network construction for symmetry-protected topological
  phases},}\ }\href {\doibase 10.1103/PhysRevB.95.125107} {\bibfield  {journal}
  {\bibinfo  {journal} {Phys. Rev. B}\ }\textbf {\bibinfo {volume} {95}},\
  \bibinfo {pages} {125107} (\bibinfo {year} {2017})}\BibitemShut {NoStop}%
\bibitem [{\citenamefont {Kruthoff}\ \emph {et~al.}(2017)\citenamefont
  {Kruthoff}, \citenamefont {de~Boer}, \citenamefont {van Wezel}, \citenamefont
  {Kane},\ and\ \citenamefont {Slager}}]{Kane2017}%
  \BibitemOpen
  \bibfield  {author} {\bibinfo {author} {\bibfnamefont {J.}~\bibnamefont
  {Kruthoff}}, \bibinfo {author} {\bibfnamefont {J.}~\bibnamefont {de~Boer}},
  \bibinfo {author} {\bibfnamefont {J.}~\bibnamefont {van Wezel}}, \bibinfo
  {author} {\bibfnamefont {C.~L.}\ \bibnamefont {Kane}}, \ and\ \bibinfo
  {author} {\bibfnamefont {R.-J.}\ \bibnamefont {Slager}},\ }\bibfield  {title}
  {\enquote {\bibinfo {title} {Topological classification of crystalline
  insulators through band structure combinatorics},}\ }\href {\doibase
  10.1103/PhysRevX.7.041069} {\bibfield  {journal} {\bibinfo  {journal} {Phys.
  Rev. X}\ }\textbf {\bibinfo {volume} {7}},\ \bibinfo {pages} {041069}
  (\bibinfo {year} {2017})}\BibitemShut {NoStop}%
\bibitem [{\citenamefont {Shiozaki}\ \emph
  {et~al.}(2018{\natexlab{a}})\citenamefont {Shiozaki}, \citenamefont {Sato},\
  and\ \citenamefont {Gomi}}]{Shiozaki2018}%
  \BibitemOpen
  \bibfield  {author} {\bibinfo {author} {\bibfnamefont {Ken}\ \bibnamefont
  {Shiozaki}}, \bibinfo {author} {\bibfnamefont {Masatoshi}\ \bibnamefont
  {Sato}}, \ and\ \bibinfo {author} {\bibfnamefont {Kiyonori}\ \bibnamefont
  {Gomi}},\ }\bibfield  {title} {\enquote {\bibinfo {title} {Atiyah-hirzebruch
  spectral sequence in band topology: General formalism and topological
  invariants for 230 space groups},}\ }\href@noop {} {\  (\bibinfo {year}
  {2018}{\natexlab{a}})},\ \Eprint {http://arxiv.org/abs/1802.06694}
  {arXiv:1802.06694 [cond-mat.str-el]} \BibitemShut {NoStop}%
\bibitem [{\citenamefont {Son{g}}\ \emph {et~al.}(2019)\citenamefont {Son{g}},
  \citenamefont {Huang}, \citenamefont {Qi}, \citenamefont {Fang},\ and\
  \citenamefont {Hermele}}]{ZDSong2018}%
  \BibitemOpen
  \bibfield  {author} {\bibinfo {author} {\bibfnamefont {Zhida}\ \bibnamefont
  {Son{g}}}, \bibinfo {author} {\bibfnamefont {Sheng-Jie}\ \bibnamefont
  {Huang}}, \bibinfo {author} {\bibfnamefont {Yang}\ \bibnamefont {Qi}},
  \bibinfo {author} {\bibfnamefont {Chen}\ \bibnamefont {Fang}}, \ and\
  \bibinfo {author} {\bibfnamefont {Michael}\ \bibnamefont {Hermele}},\
  }\bibfield  {title} {\enquote {\bibinfo {title} {Topological states from
  topological crystals},}\ }\href {\doibase 10.1126/sciadv.aax2007} {\bibfield
  {journal} {\bibinfo  {journal} {Sci. Adv.}\ }\textbf {\bibinfo {volume}
  {5}},\ \bibinfo {pages} {eaax2007} (\bibinfo {year} {2019})},\ \Eprint
  {http://arxiv.org/abs/1810.02330} {arXiv:1810.02330 [cond-mat.mes-hall]}
  \BibitemShut {NoStop}%
\bibitem [{\citenamefont {Else}\ and\ \citenamefont
  {Thorngren}(2019)}]{defect}%
  \BibitemOpen
  \bibfield  {author} {\bibinfo {author} {\bibfnamefont {D.~V.}\ \bibnamefont
  {Else}}\ and\ \bibinfo {author} {\bibfnamefont {R.}~\bibnamefont
  {Thorngren}},\ }\bibfield  {title} {\enquote {\bibinfo {title} {Crystalline
  topological phases as defect networks},}\ }\href
  {https://journals.aps.org/prb/abstract/10.1103/PhysRevB.99.115116} {\bibfield
   {journal} {\bibinfo  {journal} {Phys. Rev. B}\ }\textbf {\bibinfo {volume}
  {99}},\ \bibinfo {pages} {115116} (\bibinfo {year} {2019})}\BibitemShut
  {NoStop}%
\bibitem [{\citenamefont {Song}\ \emph
  {et~al.}(2020{\natexlab{b}})\citenamefont {Song}, \citenamefont {Fang},\ and\
  \citenamefont {Qi}}]{realspace}%
  \BibitemOpen
  \bibfield  {author} {\bibinfo {author} {\bibfnamefont {Z.}~\bibnamefont
  {Song}}, \bibinfo {author} {\bibfnamefont {C.}~\bibnamefont {Fang}}, \ and\
  \bibinfo {author} {\bibfnamefont {Y.}~\bibnamefont {Qi}},\ }\bibfield
  {title} {\enquote {\bibinfo {title} {Real-space recipes for general
  topological crystalline states},}\ }\href {\doibase
  10.1038/s41467-020-17685-5} {\bibfield  {journal} {\bibinfo  {journal}
  {Nature Communications}\ }\textbf {\bibinfo {volume} {11}},\ \bibinfo {pages}
  {4197} (\bibinfo {year} {2020}{\natexlab{b}})},\ \Eprint
  {http://arxiv.org/abs/1810.11013} {arXiv:1810.11013 [cond-mat.str-el]}
  \BibitemShut {NoStop}%
\bibitem [{\citenamefont {Shiozaki}\ \emph
  {et~al.}(2018{\natexlab{b}})\citenamefont {Shiozaki}, \citenamefont {Xiong},\
  and\ \citenamefont {Gomi}}]{KenX}%
  \BibitemOpen
  \bibfield  {author} {\bibinfo {author} {\bibfnamefont {Ken}\ \bibnamefont
  {Shiozaki}}, \bibinfo {author} {\bibfnamefont {Charles~Zhaoxi}\ \bibnamefont
  {Xiong}}, \ and\ \bibinfo {author} {\bibfnamefont {Kiyonori}\ \bibnamefont
  {Gomi}},\ }\bibfield  {title} {\enquote {\bibinfo {title} {Generalized
  homology and atiyah-hirzebruch spectral sequence in crystalline symmetry
  protected topological phenomena},}\ }\href@noop {} {\  (\bibinfo {year}
  {2018}{\natexlab{b}})},\ \Eprint {http://arxiv.org/abs/1810.00801}
  {arXiv:1810.00801 [cond-mat.str-el]} \BibitemShut {NoStop}%
\bibitem [{\citenamefont {Cheng}\ and\ \citenamefont {Wang}()}]{rotation}%
  \BibitemOpen
  \bibfield  {author} {\bibinfo {author} {\bibfnamefont {M.}~\bibnamefont
  {Cheng}}\ and\ \bibinfo {author} {\bibfnamefont {C.}~\bibnamefont {Wang}},\
  }\bibfield  {title} {\enquote {\bibinfo {title} {Rotation symmetry-protected
  topological phases of fermions},}\ }\href@noop {} {\ }\Eprint
  {http://arxiv.org/abs/1810.12308} {arXiv:1810.12308 [cond-mat.str-el]}
  \BibitemShut {NoStop}%
\bibitem [{\citenamefont {Zhang}\ \emph {et~al.}(2020)\citenamefont {Zhang},
  \citenamefont {Wang}, \citenamefont {Yang}, \citenamefont {Qi},\ and\
  \citenamefont {Gu}}]{dihedral}%
  \BibitemOpen
  \bibfield  {author} {\bibinfo {author} {\bibfnamefont {J.-H.}\ \bibnamefont
  {Zhang}}, \bibinfo {author} {\bibfnamefont {Q.-R.}\ \bibnamefont {Wang}},
  \bibinfo {author} {\bibfnamefont {S.}~\bibnamefont {Yang}}, \bibinfo {author}
  {\bibfnamefont {Y.}~\bibnamefont {Qi}}, \ and\ \bibinfo {author}
  {\bibfnamefont {Z.-C.}\ \bibnamefont {Gu}},\ }\bibfield  {title} {\enquote
  {\bibinfo {title} {Construction and classification of point-group
  symmetry-protected topological phases in two-dimensional interacting
  fermionic systems},}\ }\href {\doibase 10.1103/PhysRevB.101.100501}
  {\bibfield  {journal} {\bibinfo  {journal} {Phys. Rev. B}\ }\textbf {\bibinfo
  {volume} {101}},\ \bibinfo {pages} {100501(R)} (\bibinfo {year}
  {2020})}\BibitemShut {NoStop}%
\bibitem [{\citenamefont {Rasmussen}\ and\ \citenamefont {Lu}(2020)}]{LuX}%
  \BibitemOpen
  \bibfield  {author} {\bibinfo {author} {\bibfnamefont {Alex}\ \bibnamefont
  {Rasmussen}}\ and\ \bibinfo {author} {\bibfnamefont {Yuan-Ming}\ \bibnamefont
  {Lu}},\ }\bibfield  {title} {\enquote {\bibinfo {title} {Classification and
  construction of higher-order symmetry protected topological phases of
  interacting bosons},}\ }\href {\doibase 10.1103/PhysRevB.101.085137}
  {\bibfield  {journal} {\bibinfo  {journal} {Phys. Rev. B}\ }\textbf {\bibinfo
  {volume} {101}},\ \bibinfo {pages} {085137} (\bibinfo {year} {2020})},\
  \Eprint {http://arxiv.org/abs/1809.07325} {arXiv:1809.07325
  [cond-mat.str-el]} \BibitemShut {NoStop}%
\bibitem [{\citenamefont {Rasmussen}\ and\ \citenamefont {Lu}()}]{YMLu2018}%
  \BibitemOpen
  \bibfield  {author} {\bibinfo {author} {\bibfnamefont {A.}~\bibnamefont
  {Rasmussen}}\ and\ \bibinfo {author} {\bibfnamefont {Y.-M.}\ \bibnamefont
  {Lu}},\ }\bibfield  {title} {\enquote {\bibinfo {title} {Intrinsically
  interacting topological crystalline insulators and superconductors},}\
  }\href@noop {} {\ }\Eprint {http://arxiv.org/abs/1810.12317}
  {arXiv:1810.12317 [cond-mat.str-el]} \BibitemShut {NoStop}%
\bibitem [{\citenamefont {Cheng}(2019)}]{Cheng2018}%
  \BibitemOpen
  \bibfield  {author} {\bibinfo {author} {\bibfnamefont {M.}~\bibnamefont
  {Cheng}},\ }\bibfield  {title} {\enquote {\bibinfo {title} {Fermionic
  lieb-schultz-mattis theorems and weak symmetry-protected phases},}\ }\href
  {\doibase 10.1103/PhysRevB.99.075143} {\bibfield  {journal} {\bibinfo
  {journal} {Phys. Rev. B}\ }\textbf {\bibinfo {volume} {99}},\ \bibinfo
  {pages} {075143} (\bibinfo {year} {2019})}\BibitemShut {NoStop}%
\bibitem [{\citenamefont {Huang}\ and\ \citenamefont
  {Hermele}(2018)}]{Hermele2018}%
  \BibitemOpen
  \bibfield  {author} {\bibinfo {author} {\bibfnamefont {S.-J.}\ \bibnamefont
  {Huang}}\ and\ \bibinfo {author} {\bibfnamefont {M.}~\bibnamefont
  {Hermele}},\ }\bibfield  {title} {\enquote {\bibinfo {title} {Surface field
  theories of point group symmetry protected topological phases},}\ }\href
  {\doibase 10.1103/PhysRevB.97.075145} {\bibfield  {journal} {\bibinfo
  {journal} {Phys. Rev. B}\ }\textbf {\bibinfo {volume} {97}},\ \bibinfo
  {pages} {075145} (\bibinfo {year} {2018})}\BibitemShut {NoStop}%
\bibitem [{\citenamefont {Ono}\ \emph {et~al.}()\citenamefont {Ono},
  \citenamefont {Po},\ and\ \citenamefont {Shiozaki}}]{Po2020}%
  \BibitemOpen
  \bibfield  {author} {\bibinfo {author} {\bibfnamefont {S.}~\bibnamefont
  {Ono}}, \bibinfo {author} {\bibfnamefont {H.~C.}\ \bibnamefont {Po}}, \ and\
  \bibinfo {author} {\bibfnamefont {K.}~\bibnamefont {Shiozaki}},\ }\bibfield
  {title} {\enquote {\bibinfo {title} {$\mathbb{Z}_2$-enriched symmetry
  indicators for topological superconductors in the 1651 magnetic space
  groups},}\ }\href@noop {} {\ }\Eprint {http://arxiv.org/abs/2008.05499}
  {arXiv:2008.05499 [cond-mat.supr-con]} \BibitemShut {NoStop}%
\bibitem [{\citenamefont {Huang}(2020)}]{Huang2020PRR}%
  \BibitemOpen
  \bibfield  {author} {\bibinfo {author} {\bibfnamefont {S.-J.}\ \bibnamefont
  {Huang}},\ }\bibfield  {title} {\enquote {\bibinfo {title} {4d
  beyond-cohomology topologicalphase protected by $c_2$ symmetry and its
  boundary theories},}\ }\href {\doibase 10.1103/PhysRevResearch.2.033236}
  {\bibfield  {journal} {\bibinfo  {journal} {Phys. Rev. Research}\ }\textbf
  {\bibinfo {volume} {2}},\ \bibinfo {pages} {033236} (\bibinfo {year}
  {2020})}\BibitemShut {NoStop}%
\bibitem [{\citenamefont {Huang}\ and\ \citenamefont
  {Hsu}(2021)}]{Huang2021PRR}%
  \BibitemOpen
  \bibfield  {author} {\bibinfo {author} {\bibfnamefont {S.-J.}\ \bibnamefont
  {Huang}}\ and\ \bibinfo {author} {\bibfnamefont {Y.-T.}\ \bibnamefont
  {Hsu}},\ }\bibfield  {title} {\enquote {\bibinfo {title} {Faithful
  derivationof symmetry indicators: A case study for topological
  superconductors with time-reversal and inversion symmetries},}\ }\href
  {\doibase 10.1103/PhysRevResearch.3.013243} {\bibfield  {journal} {\bibinfo
  {journal} {Phys. Rev. Research}\ }\textbf {\bibinfo {volume} {3}},\ \bibinfo
  {pages} {013243} (\bibinfo {year} {2021})}\BibitemShut {NoStop}%
\bibitem [{\citenamefont {Zhang}\ \emph {et~al.}()\citenamefont {Zhang},
  \citenamefont {Yang}, \citenamefont {Qi},\ and\ \citenamefont
  {Gu}}]{wallpaper}%
  \BibitemOpen
  \bibfield  {author} {\bibinfo {author} {\bibfnamefont {J.-H.}\ \bibnamefont
  {Zhang}}, \bibinfo {author} {\bibfnamefont {S.}~\bibnamefont {Yang}},
  \bibinfo {author} {\bibfnamefont {Y.}~\bibnamefont {Qi}}, \ and\ \bibinfo
  {author} {\bibfnamefont {Z.-C.}\ \bibnamefont {Gu}},\ }\bibfield  {title}
  {\enquote {\bibinfo {title} {Real-space construction of crystalline
  topological superconductors and insulators in 2d interacting fermionic
  systems},}\ }\href@noop {} {\ }\Eprint {http://arxiv.org/abs/2012.15657}
  {arXiv:2012.15657 [cond-mat.str-el]} \BibitemShut {NoStop}%
\bibitem [{\citenamefont {Zhang}\ and\ \citenamefont {Y}(2021)}]{PEPS}%
  \BibitemOpen
  \bibfield  {author} {\bibinfo {author} {\bibfnamefont {Jian-Hao}\
  \bibnamefont {Zhang}}\ and\ \bibinfo {author} {\bibnamefont {Y}},\ }\bibfield
   {title} {\enquote {\bibinfo {title} {Tensor network representations of
  fermionic crystalline topological phases on two-dimensional lattices},}\
  }\href@noop {} {\  (\bibinfo {year} {2021})},\ \Eprint
  {http://arxiv.org/abs/2109.06118} {arXiv:2109.06118 [cond-mat.str-el]}
  \BibitemShut {NoStop}%
\bibitem [{\citenamefont {Manjunath}\ and\ \citenamefont
  {Barkeshli}(2021)}]{Maissam2020}%
  \BibitemOpen
  \bibfield  {author} {\bibinfo {author} {\bibfnamefont {Naren}\ \bibnamefont
  {Manjunath}}\ and\ \bibinfo {author} {\bibfnamefont {Maissam}\ \bibnamefont
  {Barkeshli}},\ }\bibfield  {title} {\enquote {\bibinfo {title} {Crystalline
  gauge fields and quantized discrete geometric response for abelian
  topological phases with lattice symmetry},}\ }\href {\doibase
  10.1103/PhysRevResearch.3.013040} {\bibfield  {journal} {\bibinfo  {journal}
  {Phys. Rev. Research}\ }\textbf {\bibinfo {volume} {3}},\ \bibinfo {pages}
  {013040} (\bibinfo {year} {2021})}\BibitemShut {NoStop}%
\bibitem [{\citenamefont {Barkeshli}\ \emph {et~al.}(2021)\citenamefont
  {Barkeshli}, \citenamefont {Chen}, \citenamefont {Hsin},\ and\ \citenamefont
  {Manjunath}}]{Maissam2021}%
  \BibitemOpen
  \bibfield  {author} {\bibinfo {author} {\bibfnamefont {Maissam}\ \bibnamefont
  {Barkeshli}}, \bibinfo {author} {\bibfnamefont {Yu-An}\ \bibnamefont {Chen}},
  \bibinfo {author} {\bibfnamefont {Po-Shen}\ \bibnamefont {Hsin}}, \ and\
  \bibinfo {author} {\bibfnamefont {Naren}\ \bibnamefont {Manjunath}},\
  }\bibfield  {title} {\enquote {\bibinfo {title} {Classification of (2+1)d
  invertible fermionic topological phases with symmetry},}\ }\href@noop {}
  {\bibfield  {journal} {\bibinfo  {journal} {arXiv e-prints}\ ,\ \bibinfo
  {eid} {arXiv:2109.11039}} (\bibinfo {year} {2021})},\ \Eprint
  {http://arxiv.org/abs/2109.11039} {arXiv:2109.11039 [cond-mat.str-el]}
  \BibitemShut {NoStop}%
\bibitem [{\citenamefont {{Fidkowski}}\ \emph {et~al.}(2018)\citenamefont
  {{Fidkowski}}, \citenamefont {{Vishwanath}},\ and\ \citenamefont
  {{Metlitski}}}]{Max18}%
  \BibitemOpen
  \bibfield  {author} {\bibinfo {author} {\bibfnamefont {Lukasz}\ \bibnamefont
  {{Fidkowski}}}, \bibinfo {author} {\bibfnamefont {Ashvin}\ \bibnamefont
  {{Vishwanath}}}, \ and\ \bibinfo {author} {\bibfnamefont {Max~A.}\
  \bibnamefont {{Metlitski}}},\ }\bibfield  {title} {\enquote {\bibinfo {title}
  {{Surface Topological Order and a new 't Hooft Anomaly of Interaction Enabled
  3+1D Fermion SPTs}},}\ }\href@noop {} {\bibfield  {journal} {\bibinfo
  {journal} {arXiv e-prints}\ ,\ \bibinfo {eid} {arXiv:1804.08628}} (\bibinfo
  {year} {2018})},\ \Eprint {http://arxiv.org/abs/1804.08628} {arXiv:1804.08628
  [cond-mat.str-el]} \BibitemShut {NoStop}%
\bibitem [{\citenamefont {{Sullivan}}\ and\ \citenamefont
  {{Cheng}}(2020)}]{JosephMeng19}%
  \BibitemOpen
  \bibfield  {author} {\bibinfo {author} {\bibfnamefont {Joseph}\ \bibnamefont
  {{Sullivan}}}\ and\ \bibinfo {author} {\bibfnamefont {Meng}\ \bibnamefont
  {{Cheng}}},\ }\bibfield  {title} {\enquote {\bibinfo {title} {{Interacting
  edge states of fermionic symmetry-protected topological phases in two
  dimensions}},}\ }\href {\doibase 10.21468/SciPostPhys.9.2.016} {\bibfield
  {journal} {\bibinfo  {journal} {SciPost Physics}\ }\textbf {\bibinfo {volume}
  {9}},\ \bibinfo {eid} {016} (\bibinfo {year} {2020})},\ \Eprint
  {http://arxiv.org/abs/1904.08953} {arXiv:1904.08953 [cond-mat.str-el]}
  \BibitemShut {NoStop}%
\bibitem [{\citenamefont {Zhang}\ \emph {et~al.}(2022)\citenamefont {Zhang},
  \citenamefont {Qi},\ and\ \citenamefont {Gu}}]{3Dpoint}%
  \BibitemOpen
  \bibfield  {author} {\bibinfo {author} {\bibfnamefont {Jian-Hao}\
  \bibnamefont {Zhang}}, \bibinfo {author} {\bibfnamefont {Yang}\ \bibnamefont
  {Qi}}, \ and\ \bibinfo {author} {\bibfnamefont {Zheng-Cheng}\ \bibnamefont
  {Gu}},\ }\bibfield  {title} {\enquote {\bibinfo {title} {Construction and
  classification of crystalline topological superconductor and insulators in
  three-dimensional interacting fermion systems},}\ }\href@noop {} {\
  (\bibinfo {year} {2022})},\ \Eprint {http://arxiv.org/abs/2204.13558}
  {arXiv:2204.13558 [cond-mat.str-el]} \BibitemShut {NoStop}%
\bibitem [{\citenamefont {Tanaka}\ \emph {et~al.}(2012)\citenamefont {Tanaka},
  \citenamefont {Ren}, \citenamefont {Sato}, \citenamefont {Nakayama},
  \citenamefont {Souma}, \citenamefont {Takahashi}, \citenamefont {Segawa},\
  and\ \citenamefont {Ando}}]{TCIrealization1}%
  \BibitemOpen
  \bibfield  {author} {\bibinfo {author} {\bibfnamefont {Y.}~\bibnamefont
  {Tanaka}}, \bibinfo {author} {\bibfnamefont {Z.}~\bibnamefont {Ren}},
  \bibinfo {author} {\bibfnamefont {T.}~\bibnamefont {Sato}}, \bibinfo {author}
  {\bibfnamefont {K.}~\bibnamefont {Nakayama}}, \bibinfo {author}
  {\bibfnamefont {S.}~\bibnamefont {Souma}}, \bibinfo {author} {\bibfnamefont
  {T.}~\bibnamefont {Takahashi}}, \bibinfo {author} {\bibfnamefont {Kouji}\
  \bibnamefont {Segawa}}, \ and\ \bibinfo {author} {\bibfnamefont {Yoichi}\
  \bibnamefont {Ando}},\ }\bibfield  {title} {\enquote {\bibinfo {title}
  {Experimental realization of a topological crystalline insulator in snte},}\
  }\href {\doibase 10.1038/nphys2442} {\bibfield  {journal} {\bibinfo
  {journal} {Nature Physics}\ }\textbf {\bibinfo {volume} {8}},\ \bibinfo
  {pages} {800} (\bibinfo {year} {2012})}\BibitemShut {NoStop}%
\bibitem [{\citenamefont {Dziawa}\ \emph {et~al.}(2012)\citenamefont {Dziawa},
  \citenamefont {Kowalski}, \citenamefont {Dybko}, \citenamefont {Buczko},
  \citenamefont {Szczerbakow}, \citenamefont {Szot}, \citenamefont
  {{\L}usakowska}, \citenamefont {Balasubramanian}, \citenamefont {Wojek},
  \citenamefont {Berntsen}, \citenamefont {Tjernberg},\ and\ \citenamefont
  {Story}}]{TCIrealization2}%
  \BibitemOpen
  \bibfield  {author} {\bibinfo {author} {\bibfnamefont {P.}~\bibnamefont
  {Dziawa}}, \bibinfo {author} {\bibfnamefont {B.~J.}\ \bibnamefont
  {Kowalski}}, \bibinfo {author} {\bibfnamefont {K.}~\bibnamefont {Dybko}},
  \bibinfo {author} {\bibfnamefont {R.}~\bibnamefont {Buczko}}, \bibinfo
  {author} {\bibfnamefont {A.}~\bibnamefont {Szczerbakow}}, \bibinfo {author}
  {\bibfnamefont {M.}~\bibnamefont {Szot}}, \bibinfo {author} {\bibfnamefont
  {E.}~\bibnamefont {{\L}usakowska}}, \bibinfo {author} {\bibfnamefont
  {T.}~\bibnamefont {Balasubramanian}}, \bibinfo {author} {\bibfnamefont
  {B.~M.}\ \bibnamefont {Wojek}}, \bibinfo {author} {\bibfnamefont {M.~H.}\
  \bibnamefont {Berntsen}}, \bibinfo {author} {\bibfnamefont {O.}~\bibnamefont
  {Tjernberg}}, \ and\ \bibinfo {author} {\bibfnamefont {T.}~\bibnamefont
  {Story}},\ }\bibfield  {title} {\enquote {\bibinfo {title} {Topological
  crystalline insulator states in pb$_{1-x}$sn$_x$se},}\ }\href {\doibase
  10.1038/nmat3449} {\bibfield  {journal} {\bibinfo  {journal} {Nature
  Materials}\ }\textbf {\bibinfo {volume} {11}},\ \bibinfo {pages} {1023}
  (\bibinfo {year} {2012})}\BibitemShut {NoStop}%
\bibitem [{\citenamefont {Okada}\ \emph {et~al.}(2013)\citenamefont {Okada},
  \citenamefont {Serbyn}, \citenamefont {Lin}, \citenamefont {Walkup},
  \citenamefont {Zhou}, \citenamefont {Dhital}, \citenamefont {Neupane},
  \citenamefont {Xu}, \citenamefont {Wang}, \citenamefont {Sankar},
  \citenamefont {Chou}, \citenamefont {Bansil}, \citenamefont {Hasan},
  \citenamefont {Wilson}, \citenamefont {Fu},\ and\ \citenamefont
  {Madhavan}}]{TCIrealization3}%
  \BibitemOpen
  \bibfield  {author} {\bibinfo {author} {\bibfnamefont {Y.}~\bibnamefont
  {Okada}}, \bibinfo {author} {\bibfnamefont {M.}~\bibnamefont {Serbyn}},
  \bibinfo {author} {\bibfnamefont {H.}~\bibnamefont {Lin}}, \bibinfo {author}
  {\bibfnamefont {D.}~\bibnamefont {Walkup}}, \bibinfo {author} {\bibfnamefont
  {W.}~\bibnamefont {Zhou}}, \bibinfo {author} {\bibfnamefont {C.}~\bibnamefont
  {Dhital}}, \bibinfo {author} {\bibfnamefont {M.}~\bibnamefont {Neupane}},
  \bibinfo {author} {\bibfnamefont {S.}~\bibnamefont {Xu}}, \bibinfo {author}
  {\bibfnamefont {Y.~J.}\ \bibnamefont {Wang}}, \bibinfo {author}
  {\bibfnamefont {R.}~\bibnamefont {Sankar}}, \bibinfo {author} {\bibfnamefont
  {F.}~\bibnamefont {Chou}}, \bibinfo {author} {\bibfnamefont {A.}~\bibnamefont
  {Bansil}}, \bibinfo {author} {\bibfnamefont {M.~Z.}\ \bibnamefont {Hasan}},
  \bibinfo {author} {\bibfnamefont {S.~D.}\ \bibnamefont {Wilson}}, \bibinfo
  {author} {\bibfnamefont {L.}~\bibnamefont {Fu}}, \ and\ \bibinfo {author}
  {\bibfnamefont {V.}~\bibnamefont {Madhavan}},\ }\bibfield  {title} {\enquote
  {\bibinfo {title} {Observation of dirac node formation and mass acquisition
  in a topological crystalline insulator},}\ }\href
  {https://science.sciencemag.org/content/341/6153/1496/tab-figures-data}
  {\bibfield  {journal} {\bibinfo  {journal} {Science}\ }\textbf {\bibinfo
  {volume} {341}},\ \bibinfo {pages} {1496} (\bibinfo {year}
  {2013})}\BibitemShut {NoStop}%
\bibitem [{\citenamefont {Ma}\ \emph {et~al.}(2017)\citenamefont {Ma},
  \citenamefont {Yi}, \citenamefont {Lv}, \citenamefont {Wang}, \citenamefont
  {Nie}, \citenamefont {Wang}, \citenamefont {Kong}, \citenamefont {Huang},
  \citenamefont {Richard}, \citenamefont {Zhang}, \citenamefont {Yaji},
  \citenamefont {Kurado}, \citenamefont {Shin}, \citenamefont {Weng},
  \citenamefont {Bernevig}, \citenamefont {Shi}, \citenamefont {Qian},\ and\
  \citenamefont {Ding}}]{TCIrealization4}%
  \BibitemOpen
  \bibfield  {author} {\bibinfo {author} {\bibfnamefont {J.}~\bibnamefont
  {Ma}}, \bibinfo {author} {\bibfnamefont {C.}~\bibnamefont {Yi}}, \bibinfo
  {author} {\bibfnamefont {B.}~\bibnamefont {Lv}}, \bibinfo {author}
  {\bibfnamefont {Z.}~\bibnamefont {Wang}}, \bibinfo {author} {\bibfnamefont
  {S.}~\bibnamefont {Nie}}, \bibinfo {author} {\bibfnamefont {L.}~\bibnamefont
  {Wang}}, \bibinfo {author} {\bibfnamefont {L.}~\bibnamefont {Kong}}, \bibinfo
  {author} {\bibfnamefont {Y.}~\bibnamefont {Huang}}, \bibinfo {author}
  {\bibfnamefont {P.}~\bibnamefont {Richard}}, \bibinfo {author} {\bibfnamefont
  {P.}~\bibnamefont {Zhang}}, \bibinfo {author} {\bibfnamefont
  {K.}~\bibnamefont {Yaji}}, \bibinfo {author} {\bibfnamefont {K.}~\bibnamefont
  {Kurado}}, \bibinfo {author} {\bibfnamefont {S.}~\bibnamefont {Shin}},
  \bibinfo {author} {\bibfnamefont {H.}~\bibnamefont {Weng}}, \bibinfo {author}
  {\bibfnamefont {B.~A.}\ \bibnamefont {Bernevig}}, \bibinfo {author}
  {\bibfnamefont {Y.}~\bibnamefont {Shi}}, \bibinfo {author} {\bibfnamefont
  {T.}~\bibnamefont {Qian}}, \ and\ \bibinfo {author} {\bibfnamefont
  {H.}~\bibnamefont {Ding}},\ }\bibfield  {title} {\enquote {\bibinfo {title}
  {Experimental evidence of hourglass fermion in the candidate nonsymmorphic
  topological insulator khgsb},}\ }\href
  {https://advances.sciencemag.org/content/3/5/e1602415} {\bibfield  {journal}
  {\bibinfo  {journal} {Sci. Adv.}\ }\textbf {\bibinfo {volume} {3}},\ \bibinfo
  {pages} {e1602415} (\bibinfo {year} {2017})}\BibitemShut {NoStop}%
\bibitem [{\citenamefont {Wang}\ \emph
  {et~al.}(2018{\natexlab{a}})\citenamefont {Wang}, \citenamefont {Liu},
  \citenamefont {Lu}, ,\ and\ \citenamefont {Zhang}}]{Wang2018}%
  \BibitemOpen
  \bibfield  {author} {\bibinfo {author} {\bibfnamefont {Qiyue}\ \bibnamefont
  {Wang}}, \bibinfo {author} {\bibfnamefont {Cheng-Cheng}\ \bibnamefont {Liu}},
  \bibinfo {author} {\bibfnamefont {Yuan-Ming}\ \bibnamefont {Lu}}, , \ and\
  \bibinfo {author} {\bibfnamefont {Fan}\ \bibnamefont {Zhang}},\ }\bibfield
  {title} {\enquote {\bibinfo {title} {High-temperature majorana corner
  states},}\ }\href {\doibase 10.1103/PhysRevLett.121.186801} {\bibfield
  {journal} {\bibinfo  {journal} {Phys. Rev. Lett.}\ }\textbf {\bibinfo
  {volume} {121}},\ \bibinfo {pages} {186801} (\bibinfo {year}
  {2018}{\natexlab{a}})}\BibitemShut {NoStop}%
\bibitem [{\citenamefont {Yan}\ \emph {et~al.}(2018)\citenamefont {Yan},
  \citenamefont {Song},\ and\ \citenamefont {Wang}}]{Yan2018}%
  \BibitemOpen
  \bibfield  {author} {\bibinfo {author} {\bibfnamefont {Zhongbo}\ \bibnamefont
  {Yan}}, \bibinfo {author} {\bibfnamefont {Fei}\ \bibnamefont {Song}}, \ and\
  \bibinfo {author} {\bibfnamefont {Zhong}\ \bibnamefont {Wang}},\ }\bibfield
  {title} {\enquote {\bibinfo {title} {Majorana corner modes in a
  high-temperature platform},}\ }\href {\doibase
  10.1103/PhysRevLett.121.096803} {\bibfield  {journal} {\bibinfo  {journal}
  {Phys. Rev. Lett.}\ }\textbf {\bibinfo {volume} {121}},\ \bibinfo {pages}
  {096803} (\bibinfo {year} {2018})}\BibitemShut {NoStop}%
\bibitem [{\citenamefont {Liu}\ \emph {et~al.}(2018)\citenamefont {Liu},
  \citenamefont {He},\ and\ \citenamefont {Nori}}]{Nori2018}%
  \BibitemOpen
  \bibfield  {author} {\bibinfo {author} {\bibfnamefont {Tao}\ \bibnamefont
  {Liu}}, \bibinfo {author} {\bibfnamefont {James~Jun}\ \bibnamefont {He}}, \
  and\ \bibinfo {author} {\bibfnamefont {Franco}\ \bibnamefont {Nori}},\
  }\bibfield  {title} {\enquote {\bibinfo {title} {Majorana corner states in a
  two-dimensional magnetic topological insulator on a high-temperature
  superconductor},}\ }\href {\doibase 10.1103/PhysRevB.98.245413} {\bibfield
  {journal} {\bibinfo  {journal} {Phys. Rev. B}\ }\textbf {\bibinfo {volume}
  {98}},\ \bibinfo {pages} {245413} (\bibinfo {year} {2018})}\BibitemShut
  {NoStop}%
\bibitem [{\citenamefont {Wang}\ \emph
  {et~al.}(2018{\natexlab{b}})\citenamefont {Wang}, \citenamefont {Lin},\ and\
  \citenamefont {Hughes}}]{Wangyuxuan2018}%
  \BibitemOpen
  \bibfield  {author} {\bibinfo {author} {\bibfnamefont {Yuxuan}\ \bibnamefont
  {Wang}}, \bibinfo {author} {\bibfnamefont {Mao}\ \bibnamefont {Lin}}, \ and\
  \bibinfo {author} {\bibfnamefont {Taylor~L.}\ \bibnamefont {Hughes}},\
  }\bibfield  {title} {\enquote {\bibinfo {title} {Weak-pairing higher order
  topological superconductors},}\ }\href {\doibase 10.1103/PhysRevB.98.165144}
  {\bibfield  {journal} {\bibinfo  {journal} {Phys. Rev. B}\ }\textbf {\bibinfo
  {volume} {98}},\ \bibinfo {pages} {165144} (\bibinfo {year}
  {2018}{\natexlab{b}})}\BibitemShut {NoStop}%
\bibitem [{\citenamefont {Shapourian}\ \emph {et~al.}(2018)\citenamefont
  {Shapourian}, \citenamefont {Wang},\ and\ \citenamefont {Ryu}}]{Ryu2018}%
  \BibitemOpen
  \bibfield  {author} {\bibinfo {author} {\bibfnamefont {Hassan}\ \bibnamefont
  {Shapourian}}, \bibinfo {author} {\bibfnamefont {Yuxuan}\ \bibnamefont
  {Wang}}, \ and\ \bibinfo {author} {\bibfnamefont {Shinsei}\ \bibnamefont
  {Ryu}},\ }\bibfield  {title} {\enquote {\bibinfo {title} {Topological
  crystalline superconductivity and second-order topological superconductivity
  in nodal-loop materials},}\ }\href {\doibase 10.1103/PhysRevB.97.094508}
  {\bibfield  {journal} {\bibinfo  {journal} {Phys. Rev. B}\ }\textbf {\bibinfo
  {volume} {97}},\ \bibinfo {pages} {094508} (\bibinfo {year}
  {2018})}\BibitemShut {NoStop}%
\bibitem [{\citenamefont {Zhang}\ \emph {et~al.}(2019)\citenamefont {Zhang},
  \citenamefont {Cole},\ and\ \citenamefont {Sarma}}]{Zhang2019}%
  \BibitemOpen
  \bibfield  {author} {\bibinfo {author} {\bibfnamefont {Rui-Xing}\
  \bibnamefont {Zhang}}, \bibinfo {author} {\bibfnamefont {William~S.}\
  \bibnamefont {Cole}}, \ and\ \bibinfo {author} {\bibfnamefont {S.~Das}\
  \bibnamefont {Sarma}},\ }\bibfield  {title} {\enquote {\bibinfo {title}
  {Helical hinge majorana modes in iron-based superconductors},}\ }\href
  {\doibase 10.1103/PhysRevLett.122.187001} {\bibfield  {journal} {\bibinfo
  {journal} {Phys. Rev. Lett.}\ }\textbf {\bibinfo {volume} {122}},\ \bibinfo
  {pages} {187001} (\bibinfo {year} {2019})}\BibitemShut {NoStop}%
\bibitem [{\citenamefont {Hsu}\ \emph {et~al.}(2018)\citenamefont {Hsu},
  \citenamefont {Stano}, \citenamefont {Klinovaja},\ and\ \citenamefont
  {Loss}}]{Hsu2018}%
  \BibitemOpen
  \bibfield  {author} {\bibinfo {author} {\bibfnamefont {Chen-Hsuan}\
  \bibnamefont {Hsu}}, \bibinfo {author} {\bibfnamefont {Peter}\ \bibnamefont
  {Stano}}, \bibinfo {author} {\bibfnamefont {Jelena}\ \bibnamefont
  {Klinovaja}}, \ and\ \bibinfo {author} {\bibfnamefont {Daniel}\ \bibnamefont
  {Loss}},\ }\bibfield  {title} {\enquote {\bibinfo {title} {Majorana kramers
  pairs in higher-order topological insulators},}\ }\href {\doibase
  10.1103/PhysRevLett.121.196801} {\bibfield  {journal} {\bibinfo  {journal}
  {Phys. Rev. Lett.}\ }\textbf {\bibinfo {volume} {121}},\ \bibinfo {pages}
  {196801} (\bibinfo {year} {2018})}\BibitemShut {NoStop}%
\bibitem [{\citenamefont {Bultinck}\ \emph {et~al.}(2019)\citenamefont
  {Bultinck}, \citenamefont {Bernevig},\ and\ \citenamefont
  {Zaletel}}]{bultinck2019three}%
  \BibitemOpen
  \bibfield  {author} {\bibinfo {author} {\bibfnamefont {N.}~\bibnamefont
  {Bultinck}}, \bibinfo {author} {\bibfnamefont {B.~A.}\ \bibnamefont
  {Bernevig}}, \ and\ \bibinfo {author} {\bibfnamefont {M.~P.}\ \bibnamefont
  {Zaletel}},\ }\bibfield  {title} {\enquote {\bibinfo {title}
  {Three-dimensional superconductors with hybrid higher-order topology},}\
  }\href {\doibase 10.1103/PhysRevB.99.125149} {\bibfield  {journal} {\bibinfo
  {journal} {Phys. Rev. B}\ }\textbf {\bibinfo {volume} {99}},\ \bibinfo
  {pages} {125149} (\bibinfo {year} {2019})}\BibitemShut {NoStop}%
\bibitem [{\citenamefont {Roy}(2020)}]{Roy_2020}%
  \BibitemOpen
  \bibfield  {author} {\bibinfo {author} {\bibfnamefont {Bitan}\ \bibnamefont
  {Roy}},\ }\bibfield  {title} {\enquote {\bibinfo {title} {Higher-order
  topological superconductors in p -, t -odd quadrupolar dirac materials},}\
  }\href {\doibase 10.1103/PhysRevB.101.220506} {\bibfield  {journal} {\bibinfo
   {journal} {Physical Review B}\ }\textbf {\bibinfo {volume} {101}},\ \bibinfo
  {pages} {220506} (\bibinfo {year} {2020})}\BibitemShut {NoStop}%
\bibitem [{\citenamefont {Roy}\ and\ \citenamefont
  {Juri{\v{c}}i{\'{c}}}(2021)}]{Roy_2021}%
  \BibitemOpen
  \bibfield  {author} {\bibinfo {author} {\bibfnamefont {Bitan}\ \bibnamefont
  {Roy}}\ and\ \bibinfo {author} {\bibfnamefont {Vladimir}\ \bibnamefont
  {Juri{\v{c}}i{\'{c}}}},\ }\bibfield  {title} {\enquote {\bibinfo {title}
  {Mixed-parity octupolar pairing and corner majorana modes in three
  dimensions},}\ }\href {\doibase 10.1103/PhysRevB.104.L180503} {\bibfield
  {journal} {\bibinfo  {journal} {Physical Review B}\ }\textbf {\bibinfo
  {volume} {104}},\ \bibinfo {pages} {l180503} (\bibinfo {year}
  {2021})}\BibitemShut {NoStop}%
\bibitem [{\citenamefont {Laubscher}\ \emph {et~al.}(2019)\citenamefont
  {Laubscher}, \citenamefont {Loss},\ and\ \citenamefont
  {Klinovaja}}]{Laubscher_2019}%
  \BibitemOpen
  \bibfield  {author} {\bibinfo {author} {\bibfnamefont {Katharina}\
  \bibnamefont {Laubscher}}, \bibinfo {author} {\bibfnamefont {Daniel}\
  \bibnamefont {Loss}}, \ and\ \bibinfo {author} {\bibfnamefont {Jelena}\
  \bibnamefont {Klinovaja}},\ }\bibfield  {title} {\enquote {\bibinfo {title}
  {Fractional topological superconductivity and parafermion corner states},}\
  }\href {\doibase 10.1103/PhysRevResearch.1.032017} {\bibfield  {journal}
  {\bibinfo  {journal} {Physical Review Research}\ }\textbf {\bibinfo {volume}
  {1}},\ \bibinfo {pages} {032017} (\bibinfo {year} {2019})}\BibitemShut
  {NoStop}%
\bibitem [{\citenamefont {Laubscher}\ \emph {et~al.}(2020)\citenamefont
  {Laubscher}, \citenamefont {Loss},\ and\ \citenamefont
  {Klinovaja}}]{Laubscher_2020}%
  \BibitemOpen
  \bibfield  {author} {\bibinfo {author} {\bibfnamefont {Katharina}\
  \bibnamefont {Laubscher}}, \bibinfo {author} {\bibfnamefont {Daniel}\
  \bibnamefont {Loss}}, \ and\ \bibinfo {author} {\bibfnamefont {Jelena}\
  \bibnamefont {Klinovaja}},\ }\bibfield  {title} {\enquote {\bibinfo {title}
  {Majorana and parafermion corner states from two coupled sheets of bilayer
  graphene},}\ }\href {\doibase 10.1103/PhysRevResearch.2.013330} {\bibfield
  {journal} {\bibinfo  {journal} {Physical Review Research}\ }\textbf {\bibinfo
  {volume} {2}},\ \bibinfo {pages} {013330} (\bibinfo {year}
  {2020})}\BibitemShut {NoStop}%
\bibitem [{\citenamefont {Bergholtz}\ and\ \citenamefont
  {Karlhede}(2008)}]{Karlhede2008}%
  \BibitemOpen
  \bibfield  {author} {\bibinfo {author} {\bibfnamefont {E.~J.}\ \bibnamefont
  {Bergholtz}}\ and\ \bibinfo {author} {\bibfnamefont {A.}~\bibnamefont
  {Karlhede}},\ }\bibfield  {title} {\enquote {\bibinfo {title} {Quantum hall
  system in tao-thouless limit},}\ }\href {\doibase 10.1103/PhysRevB.77.155308}
  {\bibfield  {journal} {\bibinfo  {journal} {Phys. Rev. B}\ }\textbf {\bibinfo
  {volume} {77}},\ \bibinfo {pages} {155308} (\bibinfo {year}
  {2008})}\BibitemShut {NoStop}%
\bibitem [{\citenamefont {Kane}\ \emph {et~al.}(2002)\citenamefont {Kane},
  \citenamefont {Mukhopadhyay},\ and\ \citenamefont {Lubensky}}]{Lubensky2002}%
  \BibitemOpen
  \bibfield  {author} {\bibinfo {author} {\bibfnamefont {C.~L.}\ \bibnamefont
  {Kane}}, \bibinfo {author} {\bibfnamefont {Ranjan}\ \bibnamefont
  {Mukhopadhyay}}, \ and\ \bibinfo {author} {\bibfnamefont {T.~C.}\
  \bibnamefont {Lubensky}},\ }\bibfield  {title} {\enquote {\bibinfo {title}
  {Fractional quantum hall effect in an array of quantum wires},}\ }\href
  {\doibase 10.1103/PhysRevLett.88.036401} {\bibfield  {journal} {\bibinfo
  {journal} {Phys. Rev. Lett.}\ }\textbf {\bibinfo {volume} {88}},\ \bibinfo
  {pages} {036401} (\bibinfo {year} {2002})}\BibitemShut {NoStop}%
\bibitem [{\citenamefont {Teo}\ and\ \citenamefont {Kane}(2014)}]{Kane2014}%
  \BibitemOpen
  \bibfield  {author} {\bibinfo {author} {\bibfnamefont {Jeffrey C.~Y.}\
  \bibnamefont {Teo}}\ and\ \bibinfo {author} {\bibfnamefont {C.~L.}\
  \bibnamefont {Kane}},\ }\bibfield  {title} {\enquote {\bibinfo {title} {From
  luttinger liquid to non-abelian quantum hall states},}\ }\href {\doibase
  10.1103/PhysRevB.89.085101} {\bibfield  {journal} {\bibinfo  {journal} {Phys.
  Rev. B}\ }\textbf {\bibinfo {volume} {89}},\ \bibinfo {pages} {085101}
  (\bibinfo {year} {2014})}\BibitemShut {NoStop}%
\bibitem [{\citenamefont {Fuji}\ and\ \citenamefont
  {Lecheminant}(2017)}]{Lecheminant2017}%
  \BibitemOpen
  \bibfield  {author} {\bibinfo {author} {\bibfnamefont {Y.}~\bibnamefont
  {Fuji}}\ and\ \bibinfo {author} {\bibfnamefont {P.}~\bibnamefont
  {Lecheminant}},\ }\bibfield  {title} {\enquote {\bibinfo {title} {Non-abelian
  su(n-1)-singlet fractional quantum hall states from coupled wires},}\ }\href
  {\doibase 10.1103/PhysRevB.95.125130} {\bibfield  {journal} {\bibinfo
  {journal} {Phys. Rev. B}\ }\textbf {\bibinfo {volume} {95}},\ \bibinfo
  {pages} {125130} (\bibinfo {year} {2017})}\BibitemShut {NoStop}%
\bibitem [{\citenamefont {Meng}\ \emph {et~al.}(2015)\citenamefont {Meng},
  \citenamefont {Neupert}, \citenamefont {Greiter},\ and\ \citenamefont
  {Thomale}}]{Thomale2015}%
  \BibitemOpen
  \bibfield  {author} {\bibinfo {author} {\bibfnamefont {Tobias}\ \bibnamefont
  {Meng}}, \bibinfo {author} {\bibfnamefont {Titus}\ \bibnamefont {Neupert}},
  \bibinfo {author} {\bibfnamefont {Martin}\ \bibnamefont {Greiter}}, \ and\
  \bibinfo {author} {\bibfnamefont {Ronny}\ \bibnamefont {Thomale}},\
  }\bibfield  {title} {\enquote {\bibinfo {title} {Coupled-wire construction of
  chiral spin liquids},}\ }\href {\doibase 10.1103/PhysRevB.91.241106}
  {\bibfield  {journal} {\bibinfo  {journal} {Phys. Rev. B}\ }\textbf {\bibinfo
  {volume} {91}},\ \bibinfo {pages} {241106} (\bibinfo {year}
  {2015})}\BibitemShut {NoStop}%
\bibitem [{\citenamefont {Sagi}\ and\ \citenamefont {Oreg}(2015)}]{Sagi_2015}%
  \BibitemOpen
  \bibfield  {author} {\bibinfo {author} {\bibfnamefont {Eran}\ \bibnamefont
  {Sagi}}\ and\ \bibinfo {author} {\bibfnamefont {Yuval}\ \bibnamefont
  {Oreg}},\ }\bibfield  {title} {\enquote {\bibinfo {title} {From an array of
  quantum wires to three-dimensional fractional topological insulators},}\
  }\href {\doibase 10.1103/PhysRevB.92.195137} {\bibfield  {journal} {\bibinfo
  {journal} {Phys. Rev. B}\ }\textbf {\bibinfo {volume} {92}},\ \bibinfo
  {pages} {195137} (\bibinfo {year} {2015})}\BibitemShut {NoStop}%
\bibitem [{\citenamefont {Seroussi}\ \emph {et~al.}(2014)\citenamefont
  {Seroussi}, \citenamefont {Berg},\ and\ \citenamefont
  {Oreg}}]{Seroussi_2014}%
  \BibitemOpen
  \bibfield  {author} {\bibinfo {author} {\bibfnamefont {Inbar}\ \bibnamefont
  {Seroussi}}, \bibinfo {author} {\bibfnamefont {Erez}\ \bibnamefont {Berg}}, \
  and\ \bibinfo {author} {\bibfnamefont {Yuval}\ \bibnamefont {Oreg}},\
  }\bibfield  {title} {\enquote {\bibinfo {title} {Topological superconducting
  phases of weakly coupled quantum wires},}\ }\href {\doibase
  10.1103/PhysRevB.89.104523} {\bibfield  {journal} {\bibinfo  {journal} {Phys.
  Rev. B}\ }\textbf {\bibinfo {volume} {89}},\ \bibinfo {pages} {104523}
  (\bibinfo {year} {2014})}\BibitemShut {NoStop}%
\bibitem [{\citenamefont {Iadecola}\ \emph {et~al.}(2016)\citenamefont
  {Iadecola}, \citenamefont {Neupert}, \citenamefont {Chamon},\ and\
  \citenamefont {Mudry}}]{Iadecola_2016}%
  \BibitemOpen
  \bibfield  {author} {\bibinfo {author} {\bibfnamefont {Thomas}\ \bibnamefont
  {Iadecola}}, \bibinfo {author} {\bibfnamefont {Titus}\ \bibnamefont
  {Neupert}}, \bibinfo {author} {\bibfnamefont {Claudio}\ \bibnamefont
  {Chamon}}, \ and\ \bibinfo {author} {\bibfnamefont {Christopher}\
  \bibnamefont {Mudry}},\ }\bibfield  {title} {\enquote {\bibinfo {title} {Wire
  constructions of abelian topological phases in three or more dimensions},}\
  }\href {\doibase 10.1103/PhysRevB.93.195136} {\bibfield  {journal} {\bibinfo
  {journal} {Phys. Rev. B}\ }\textbf {\bibinfo {volume} {93}},\ \bibinfo
  {pages} {195136} (\bibinfo {year} {2016})}\BibitemShut {NoStop}%
\bibitem [{\citenamefont {Neupert}\ \emph {et~al.}(2014)\citenamefont
  {Neupert}, \citenamefont {Chamon}, \citenamefont {Mudry},\ and\ \citenamefont
  {Thomale}}]{Neupert_2014}%
  \BibitemOpen
  \bibfield  {author} {\bibinfo {author} {\bibfnamefont {Titus}\ \bibnamefont
  {Neupert}}, \bibinfo {author} {\bibfnamefont {Claudio}\ \bibnamefont
  {Chamon}}, \bibinfo {author} {\bibfnamefont {Christopher}\ \bibnamefont
  {Mudry}}, \ and\ \bibinfo {author} {\bibfnamefont {Ronny}\ \bibnamefont
  {Thomale}},\ }\bibfield  {title} {\enquote {\bibinfo {title} {Wire
  deconstructionism of two-dimensional topological phases},}\ }\href {\doibase
  10.1103/PhysRevB.90.205101} {\bibfield  {journal} {\bibinfo  {journal} {Phys.
  Rev. B}\ }\textbf {\bibinfo {volume} {90}},\ \bibinfo {pages} {205101}
  (\bibinfo {year} {2014})}\BibitemShut {NoStop}%
\bibitem [{\citenamefont {Klinovaja}\ and\ \citenamefont
  {Tserkovnyak}(2014)}]{Klinovaja_2014}%
  \BibitemOpen
  \bibfield  {author} {\bibinfo {author} {\bibfnamefont {Jelena}\ \bibnamefont
  {Klinovaja}}\ and\ \bibinfo {author} {\bibfnamefont {Yaroslav}\ \bibnamefont
  {Tserkovnyak}},\ }\bibfield  {title} {\enquote {\bibinfo {title} {Quantum
  spin hall effect in strip of stripes model},}\ }\href {\doibase
  10.1103/PhysRevB.90.115426} {\bibfield  {journal} {\bibinfo  {journal}
  {Physical Review B}\ }\textbf {\bibinfo {volume} {90}},\ \bibinfo {pages}
  {115426} (\bibinfo {year} {2014})}\BibitemShut {NoStop}%
\bibitem [{\citenamefont {Klinovaja}\ \emph {et~al.}(2015)\citenamefont
  {Klinovaja}, \citenamefont {Tserkovnyak},\ and\ \citenamefont
  {Loss}}]{Klinovaja_2015}%
  \BibitemOpen
  \bibfield  {author} {\bibinfo {author} {\bibfnamefont {Jelena}\ \bibnamefont
  {Klinovaja}}, \bibinfo {author} {\bibfnamefont {Yaroslav}\ \bibnamefont
  {Tserkovnyak}}, \ and\ \bibinfo {author} {\bibfnamefont {Daniel}\
  \bibnamefont {Loss}},\ }\bibfield  {title} {\enquote {\bibinfo {title}
  {Integer and fractional quantum anomalous hall effect in a strip of stripes
  model},}\ }\href {\doibase 10.1103/PhysRevB.91.085426} {\bibfield  {journal}
  {\bibinfo  {journal} {Physical Review B}\ }\textbf {\bibinfo {volume} {91}},\
  \bibinfo {pages} {085426} (\bibinfo {year} {2015})}\BibitemShut {NoStop}%
\bibitem [{\citenamefont {Klinovaja}\ and\ \citenamefont
  {Loss}(2014)}]{Klinovaja_2014a}%
  \BibitemOpen
  \bibfield  {author} {\bibinfo {author} {\bibfnamefont {Jelena}\ \bibnamefont
  {Klinovaja}}\ and\ \bibinfo {author} {\bibfnamefont {Daniel}\ \bibnamefont
  {Loss}},\ }\bibfield  {title} {\enquote {\bibinfo {title} {Integer and
  fractional quantum hall effect in a strip of stripes},}\ }\href {\doibase
  10.1140/epjb/e2014-50395-6} {\bibfield  {journal} {\bibinfo  {journal} {The
  European Physical Journal B}\ }\textbf {\bibinfo {volume} {87}} (\bibinfo
  {year} {2014}),\ 10.1140/epjb/e2014-50395-6}\BibitemShut {NoStop}%
\bibitem [{\citenamefont {Zhang}\ and\ \citenamefont {Ning}(2021)}]{BBC}%
  \BibitemOpen
  \bibfield  {author} {\bibinfo {author} {\bibfnamefont {Jian-Hao}\
  \bibnamefont {Zhang}}\ and\ \bibinfo {author} {\bibfnamefont {Shang-Qiang}\
  \bibnamefont {Ning}},\ }\bibfield  {title} {\enquote {\bibinfo {title}
  {Crystalline equivalent boundary-bulk correspondence of two-dimensional
  topological phases},}\ }\href@noop {} {\  (\bibinfo {year} {2021})},\ \Eprint
  {http://arxiv.org/abs/2112.14567} {arXiv:2112.14567 [cond-mat.str-el]}
  \BibitemShut {NoStop}%
\bibitem [{sup()}]{supplementary}%
  \BibitemOpen
  \href@noop {} {\bibinfo  {journal} {see Supplementary Materials for more
  details}\ }\BibitemShut {NoStop}%
\bibitem [{\citenamefont {Wu}\ \emph {et~al.}(2019)\citenamefont {Wu},
  \citenamefont {Jian},\ and\ \citenamefont {Xu}}]{Wu_2019}%
  \BibitemOpen
\bibfield  {journal} {  }\bibfield  {author} {\bibinfo {author} {\bibfnamefont
  {Xiao-Chuan}\ \bibnamefont {Wu}}, \bibinfo {author} {\bibfnamefont
  {Chao-Ming}\ \bibnamefont {Jian}}, \ and\ \bibinfo {author} {\bibfnamefont
  {Cenke}\ \bibnamefont {Xu}},\ }\bibfield  {title} {\enquote {\bibinfo {title}
  {Coupled-wire description of the correlated physics in twisted bilayer
  graphene},}\ }\href {\doibase 10.1103/PhysRevB.99.161405} {\bibfield
  {journal} {\bibinfo  {journal} {Physical Review B}\ }\textbf {\bibinfo
  {volume} {99}},\ \bibinfo {pages} {161405} (\bibinfo {year}
  {2019})}\BibitemShut {NoStop}%
\bibitem [{\citenamefont {Wang}\ \emph {et~al.}(2022)\citenamefont {Wang},
  \citenamefont {Yu}, \citenamefont {Kwan}, \citenamefont {Jia}, \citenamefont
  {Lei}, \citenamefont {Klemenz}, \citenamefont {Cevallos}, \citenamefont
  {Singha}, \citenamefont {Devakul}, \citenamefont {Watanabe}, \citenamefont
  {Taniguchi}, \citenamefont {Sondhi}, \citenamefont {Cava}, \citenamefont
  {Schoop}, \citenamefont {Parameswaran},\ and\ \citenamefont
  {Wu}}]{Wang_2022}%
  \BibitemOpen
  \bibfield  {author} {\bibinfo {author} {\bibfnamefont {Pengjie}\ \bibnamefont
  {Wang}}, \bibinfo {author} {\bibfnamefont {Guo}\ \bibnamefont {Yu}}, \bibinfo
  {author} {\bibfnamefont {Yves~H.}\ \bibnamefont {Kwan}}, \bibinfo {author}
  {\bibfnamefont {Yanyu}\ \bibnamefont {Jia}}, \bibinfo {author} {\bibfnamefont
  {Shiming}\ \bibnamefont {Lei}}, \bibinfo {author} {\bibfnamefont {Sebastian}\
  \bibnamefont {Klemenz}}, \bibinfo {author} {\bibfnamefont {F.~Alexandre}\
  \bibnamefont {Cevallos}}, \bibinfo {author} {\bibfnamefont {Ratnadwip}\
  \bibnamefont {Singha}}, \bibinfo {author} {\bibfnamefont {Trithep}\
  \bibnamefont {Devakul}}, \bibinfo {author} {\bibfnamefont {Kenji}\
  \bibnamefont {Watanabe}}, \bibinfo {author} {\bibfnamefont {Takashi}\
  \bibnamefont {Taniguchi}}, \bibinfo {author} {\bibfnamefont {Shivaji~L.}\
  \bibnamefont {Sondhi}}, \bibinfo {author} {\bibfnamefont {Robert~J.}\
  \bibnamefont {Cava}}, \bibinfo {author} {\bibfnamefont {Leslie~M.}\
  \bibnamefont {Schoop}}, \bibinfo {author} {\bibfnamefont {Siddharth~A.}\
  \bibnamefont {Parameswaran}}, \ and\ \bibinfo {author} {\bibfnamefont
  {Sanfeng}\ \bibnamefont {Wu}},\ }\bibfield  {title} {\enquote {\bibinfo
  {title} {One-dimensional luttinger liquids in a two-dimensional moir{\'{e}}
  lattice},}\ }\href {\doibase 10.1038/s41586-022-04514-6} {\bibfield
  {journal} {\bibinfo  {journal} {Nature}\ }\textbf {\bibinfo {volume} {605}},\
  \bibinfo {pages} {57--62} (\bibinfo {year} {2022})}\BibitemShut {NoStop}%
\end{thebibliography}

\begin{thebibliography}{3}%
\makeatletter
\providecommand \@ifxundefined [1]{%
 \@ifx{#1\undefined}
}%
\providecommand \@ifnum [1]{%
 \ifnum #1\expandafter \@firstoftwo
 \else \expandafter \@secondoftwo
 \fi
}%
\providecommand \@ifx [1]{%
 \ifx #1\expandafter \@firstoftwo
 \else \expandafter \@secondoftwo
 \fi
}%
\providecommand \natexlab [1]{#1}%
\providecommand \enquote  [1]{``#1''}%
\providecommand \bibnamefont  [1]{#1}%
\providecommand \bibfnamefont [1]{#1}%
\providecommand \citenamefont [1]{#1}%
\providecommand \href@noop [0]{\@secondoftwo}%
\providecommand \href [0]{\begingroup \@sanitize@url \@href}%
\providecommand \@href[1]{\@@startlink{#1}\@@href}%
\providecommand \@@href[1]{\endgroup#1\@@endlink}%
\providecommand \@sanitize@url [0]{\catcode `\\12\catcode `\$12\catcode
  `\&12\catcode `\#12\catcode `\^12\catcode `\_12\catcode `\%12\relax}%
\providecommand \@@startlink[1]{}%
\providecommand \@@endlink[0]{}%
\providecommand \url  [0]{\begingroup\@sanitize@url \@url }%
\providecommand \@url [1]{\endgroup\@href {#1}{\urlprefix }}%
\providecommand \urlprefix  [0]{URL }%
\providecommand \Eprint [0]{\href }%
\providecommand \doibase [0]{http://dx.doi.org/}%
\providecommand \selectlanguage [0]{\@gobble}%
\providecommand \bibinfo  [0]{\@secondoftwo}%
\providecommand \bibfield  [0]{\@secondoftwo}%
\providecommand \translation [1]{[#1]}%
\providecommand \BibitemOpen [0]{}%
\providecommand \bibitemStop [0]{}%
\providecommand \bibitemNoStop [0]{.\EOS\space}%
\providecommand \EOS [0]{\spacefactor3000\relax}%
\providecommand \BibitemShut  [1]{\csname bibitem#1\endcsname}%
\let\auto@bib@innerbib\@empty
\bibitem [{\citenamefont {Lu}\ and\ \citenamefont {Vishwanath}(2012)}]{Lu12S}%
  \BibitemOpen
  \bibfield  {author} {\bibinfo {author} {\bibfnamefont {Yuan-Ming}\
  \bibnamefont {Lu}}\ and\ \bibinfo {author} {\bibfnamefont {Ashvin}\
  \bibnamefont {Vishwanath}},\ }\bibfield  {title} {\enquote {\bibinfo {title}
  {Theory and classification of interacting integer topological phases in two
  dimensions: A chern-simons approach},}\ }\href {\doibase
  10.1103/PhysRevB.86.125119} {\bibfield  {journal} {\bibinfo  {journal} {Phys.
  Rev. B}\ }\textbf {\bibinfo {volume} {86}},\ \bibinfo {pages} {125119}
  (\bibinfo {year} {2012})}\BibitemShut {NoStop}%
\bibitem [{\citenamefont {Ning}\ \emph {et~al.}(2021)\citenamefont {Ning},
  \citenamefont {Wang}, \citenamefont {Wang},\ and\ \citenamefont
  {Gu}}]{Ning21aS}%
  \BibitemOpen
  \bibfield  {author} {\bibinfo {author} {\bibfnamefont {Shang-Qiang}\
  \bibnamefont {Ning}}, \bibinfo {author} {\bibfnamefont {Chenjie}\
  \bibnamefont {Wang}}, \bibinfo {author} {\bibfnamefont {Qing-Rui}\
  \bibnamefont {Wang}}, \ and\ \bibinfo {author} {\bibfnamefont {Zheng-Cheng}\
  \bibnamefont {Gu}},\ }\bibfield  {title} {\enquote {\bibinfo {title} {Edge
  theories of two-dimensional fermionic symmetry protected topological phases
  protected by unitary abelian symmetries},}\ }\href {\doibase
  10.1103/PhysRevB.104.075151} {\bibfield  {journal} {\bibinfo  {journal}
  {Phys. Rev. B}\ }\textbf {\bibinfo {volume} {104}},\ \bibinfo {pages}
  {075151} (\bibinfo {year} {2021})}\BibitemShut {NoStop}%
\bibitem [{\citenamefont {Haldane}(1995)}]{Haldane1995S}%
  \BibitemOpen
  \bibfield  {author} {\bibinfo {author} {\bibfnamefont {F.~D.~M.}\
  \bibnamefont {Haldane}},\ }\bibfield  {title} {\enquote {\bibinfo {title}
  {Stability of chiral luttinger liquids and abelian quantum hall states},}\
  }\href {\doibase 10.1103/PhysRevLett.74.2090} {\bibfield  {journal} {\bibinfo
   {journal} {Phys. Rev. Lett.}\ }\textbf {\bibinfo {volume} {74}},\ \bibinfo
  {pages} {2090} (\bibinfo {year} {1995})}\BibitemShut {NoStop}%
\end{thebibliography}

\pagebreak

\clearpage

\appendix
\setcounter{equation}{0}
\newpage

\renewcommand{\thesection}{S-\arabic{section}} \renewcommand{\theequation}{S%
\arabic{equation}} \setcounter{equation}{0} \renewcommand{\thefigure}{S%
\arabic{figure}} \setcounter{figure}{0}

\onecolumngrid

\vskip0.2cm
\centerline{\large\textbf{Supplemental Materials of ``Strongly correlated crystalline higher-order topological}}
\vskip0.12cm
\centerline{\large\textbf{phases in two-dimensional systems: A coupled-wire study''}}

\vskip0.8cm
\twocolumngrid

\maketitle

\section{$K$-matrix formalism of fSPT}
In the main text, we define the 1D quantum wire based on the nonchiral Luttinger liquid with topological $K$-matrix. In this section we review the $K$-matrix formalism of fermionic symmetry-protected topological (fSPT) phases. A $U(1)$ Chern-Simons theory has the form:
\begin{align}
\mathcal{L}=\frac{K_{IJ}}{4\pi}\epsilon^{\mu\nu\lambda}a_\mu^I\partial_\nu a_\lambda^J+a_\mu^Ij_I^\mu+\cdot\cdot\cdot
\label{Chern-Simons}
\end{align}
where $K$ is a symmetric integral matrix, $\{a^I\}$ is a set of one-form gauge fields, and $\{j_I\}$ are the corresponding currents that couple to the gauge fields $a^I$. The symmetry is defined as: two theories $\mathcal{L}[a^I]$ and $\mathcal{L}[\tilde{a}^I]$ correspond to the same phase if there is an $n\times n$ integral unimodular matrix $W$ satisfying $\tilde{a}^I=W_{IJ}a^J$. 

The topological order described by Abelian Chern-Simons theory hosts Abelian anyon excitations. An anyon is labeled by an integer vector $l=(l_1,l_2,\cdot\cdot\cdot,l_n)^T$. The self and mutual statistics of anyons are:
\begin{align}
\begin{aligned}
\theta_l&=\pi l^TK^{-1}l\\
\theta_{l,l'}&=2\pi l^T K^{-1}l'
\end{aligned}
\end{align}

The total number of anyons and the ground-state degeneracy (GSD) on a torus are both given by $|\mathrm{det}K|$. For SPT phase, there is no GSD or anyon, hence we require $|\mathrm{det}K|=1$ for SPT phases. 

The $K$-matrix Chern-Simons theory has a well-known bulk-boundary correspondence \cite{Lu12S,Ning21aS}. In a system with open boundary, the edge thoery of (\ref{Chern-Simons}) has the form:
\begin{align}
\mathcal{L}_{\mathrm{edge}}=\frac{K_{IJ}}{4\pi}\left(\partial_x\phi^I\right)\left(\partial_t\phi^J\right)+\frac{V_{IJ}}{8\pi}\left(\partial_x\phi^I\right)\left(\partial_x\phi^J\right)
\label{LuttingerS}
\end{align}
where $\phi=(\{\phi^I\})^T$ are chiral bosonic fields on the edge and related to dynamical gauge field $a_\mu^I$ in the bulk by $a_\mu^I=\partial_\mu\phi^I$, and an anyon on the edge can be created by the operator $e^{il^T\phi}$.

\section{Assemble of quantum wires of $C_4$-symmetric class-BD\1 HOTSC}
In the main text, the inter-wire couplings of 2D $C_4$-symmetric class-BD\1 HOTSC are defined by backscatterings of four 1D quantum wires as an assembly. In this section we demonstrate that the minimal number of quantum wires of an assembly should be four, equivalently, two quantum wires cannot be gapped in a symmetric way. 

The 1D quantum wire building block for coupled-wire construction of 2D $C_4$-symmetric class-BD\1 HOTSC is described by 1D nonchiral Luttinger liquid on a circle, with the Lagrangian: 
\begin{align}
\mathcal{L}_0^j=\frac{K_{IJ}^j}{4\pi}\left(\partial_\theta\phi^I_j\right)\left(\partial_t\phi^J_j\right)+\frac{V_{IJ}^j}{8\pi}\left(\partial_\theta\phi^I_j\right)\left(\partial_\theta\phi^J_j\right)
\label{wireS}
\end{align}
where $\theta$ is the polar angle of the circle, $\phi^j(\theta)=(\phi_1^j(\theta),\phi_2^j(\theta),\phi_3^j(\theta),\phi_4^j(\theta))^T$ as 4-component bosonic fields of $j^{\mathrm{th}}$ quantum wire, and $K^j=\sigma^z\oplus\sigma^z$ as the topological $K$-matrix. In the main text, the $(C_4\times\mathbb{Z}_2^T)$ symmetry has been defined on the bosonic fields as Eqs. ({\color{red}6}) and ({\color{red}7}), which can be reformulated to:
\begin{align}
\bs{R}:\phi^j\mapsto W^{\bs{R}}_j\phi^j+\delta\phi^{\bs{R}}_j,~~~\mathcal{T}:\phi^j\mapsto W^{\mathcal{T}}_j\phi+\delta\phi^{\mathcal{T}}_j
\end{align}
where ($\Theta$ is the operator that transforms the polar angle $\theta\mapsto\theta+\pi/2$)
\begin{align}
W^{\bs{R}}_j=\left(
\begin{array}{cccc}
1 & 0 & 1 & -1\\
0 & 1 & -1 & 1\\
1 & 1 & -1 & 0\\
1 & 1 & 0 & -1
\end{array}
\right)\Theta,~~~\delta\phi^{\bs{R}}_j=\frac{\pi}{2}\left(
\begin{array}{cccc}
-1\\
1\\
1\\
1
\end{array}
\right)
\label{BDI-RS}
\end{align}
and
\begin{align}
W^{\mathcal{T}}_j=\left(
\begin{array}{cccc}
0 & 1 & -1 & 1\\
1 & 0 & 1 & -1\\
1 & 1 & 0 & -1\\
1 & 1 & -1 & 0
\end{array}
\right),~~~\delta\phi^{\mathcal{T}}_j=\left(
\begin{array}{cccc}
0\\
\pi\\
\pi\\
0
\end{array}
\right)
\label{BDI-TS}
\end{align}

Now we consider two copies of such quantum wire and investigate whether they can be fully gapped out. The corresponding Lagrangian has the similar form to Eq. (\ref{wireS}), with $\phi(\theta)=(\phi^1(\theta),\phi^2(\theta))^T$ as 8-component bosonic fields, and $K=(\sigma^z)^{\oplus4}$ as the topological $K$-matrix. For this case, the $(C_4\times\mathbb{Z}_2^T)$ symmetry is defined on $\phi$ as:
\begin{align}
\bs{R}:\phi\mapsto W^{\bs{R}}\phi+\delta\phi^{\bs{R}},~~~\mathcal{T}:\phi\mapsto W^{\mathcal{T}}\phi+\delta\phi^{\mathcal{T}}
\end{align}
where
\begin{align}
W^{\bs{R}}=\left(
\begin{array}{cccc}
1 & 0 & 1 & -1\\
0 & 1 & -1 & 1\\
1 & 1 & -1 & 0\\
1 & 1 & 0 & -1
\end{array}
\right)\otimes\mathbbm{1}_{2\times2}\Theta
\end{align}
$\delta\phi^{\bs{R}}=\delta\phi^{\bs{R}}_1\oplus\delta\phi^{\bs{R}}_2$, $\delta\phi^{\mathcal{T}}=\delta\phi^{\mathcal{T}}_1\oplus\delta\phi^{\mathcal{T}}_2$, and
\begin{align}
W^{\mathcal{T}}=\left(
\begin{array}{cccc}
0 & 1 & -1 & 1\\
1 & 0 & 1 & -1\\
1 & 1 & 0 & -1\\
1 & 1 & -1 & 0
\end{array}
\right)\otimes\mathbbm{1}_{2\times2}
\end{align}

We now try to construct interaction terms that gap out the edge without breaking $\bs{R}$ and $\mathcal{T}$ symmetries, neither explicitly nor spontaneously. Consider the backscattering term of the form:
\begin{align}
U=\sum\limits_{k}U(\Lambda_k)=\sum\limits_{k}U(\theta)\cos\left[\Lambda_k^TK\phi-\alpha(\theta)\right]
\label{backscatteringS}
\end{align}
The backscattering term (\ref{backscatteringS}) can gap out the edge as long as the vectors $\{\Lambda_k\}$ satisfy the ``null-vector'' conditions \cite{Haldane1995S} for $\forall i,j$:
\begin{align}
\Lambda_i^TK\Lambda_j=0
\label{null-vector}
\end{align}
For the present case, there are only two linear independent solutions to this problem:
\begin{align}
\begin{aligned}
&\Lambda_1^T=\left(1,0,0,1,1,0,0,1\right)^T\\
&\Lambda_2^T=\left(0,1,1,0,0,1,1,0\right)^T
\end{aligned}
\label{solutionS}
\end{align}
Nevertheless, there are eight bosonic fields $\phi^{1,2}$, we should introduce at least 4 independent backscattering terms in order to fully gap them out, hence we cannot fully gap out the two copies of 1D quantum wires for the cases of 2D $C_4$-symmtric class-BD\1 HOTSC. 

For the case with 4 copies of 1D quantum wires, we have introduced eight linear independent 4-component backscattering terms in the main text that can fully gap all four quantum wires. As the consequence, in the coupled-wire construction of 2D $C_4$-symmetric class-BD\1 HOTSCs, we should assemble the 1D quantum wires by four, and the corresponding classification from the coupled-wire constructions is $\mathbb{Z}_4$. 

\section{Symmetry and couplings of quantum wires}
In this section, we explain the symmetry defined in the main text in details and demonstrate the inter-wire and intra-wire couplings are symmetric under arbitrary crystalline symmetry operations.

\subsection{$D_4$-symmetric class-$D$ HOTSC}
In the main text, we have defined the $D_4$-symmetry in Eq. ({\color{red}2}). Here we justify the couplings introduced in the main text are $D_4$-symmetric. 

Firstly, it is easy to verify that the $D_4$ symmetry operations defined in Eq. ({\color{red}2}) satisfy the group structure of 4-fold dihedral symmetry. Firstly, the rotation and reflection generators, $\bs{R}$ and $\bs{M}$ satisfy the following condition for spinless fermions:
\begin{align}
\bs{R}^4=\bs{M}^2=\mathbbm{1}
\end{align}
and the symmetry operation $\bs{MRM}$ will transform the bosonic fields $\phi_{1,2}^j$ as:
\begin{align}
\begin{aligned}
&\phi_1^j(\theta)\mapsto-\phi_1^j(\theta+\frac{3\pi}{2})\\
&\phi_2^j(\theta)\mapsto-\phi_2^j(\theta+\frac{3\pi}{2})+\pi
\end{aligned}
\end{align}
which is identical to $\bs{R}^3$. We can straightforwardly verify that the backscattering terms [cf. Eqs. ({\color{red}3})-({\color{red}5}) in the main text] are $D_4$-symmetric. 

Then we focus on the inter-wire and intra-wire backscattering terms and demonstrate that they can fully gap out the bulk of the $D_4$-symmetric class-$D$ HOTSC. Consider the $(2j-2+k)^{\mathrm{th}}$ and $(2j-1+k)^{\mathrm{th}}$ quantum wires, there are 4 independent bosonic fields $\phi_{1,2}^{2j-2+k}$ and $\phi_{1,2}^{2j-1+k}$. In order to gap them out, we need to introduce at least two linear independent backscattering terms satisfying the Haldane's ``null-vector'' condition [cf. Eq. (\ref{null-vector})]. It is straightforwardly to verify that the inter-wire coupling $\mathcal{L}_{ck}^j$ includes two linear independent backscattering terms satisfying the Haldane's ``null-vector'' condition, hence the related $(2j-2+k)^{\mathrm{th}}$ and $(2j-1+k)^{\mathrm{th}}$ quantum wires are fully gapped out by $\mathcal{L}_{ck}^j$.

For nontrivial HOTSC dominated by $\mathcal{L}_{c2}$, the inner-most quantum wire remains gapless at each pole. The neighborhoods of all four poles includes four segments of bosonic field, $\phi_{1,2}(\theta)$, where $\theta\in(-\epsilon+{\beta\pi}/{2},\epsilon+{\beta\pi}/{2})$, $\beta=0,1,2,3$. The intra-wire coupling $\mathcal{L}_{\mathrm{int}}$ includes only two linear independent backscattering terms globally, nonetheless, if we only focus on the gapless segments at poles of the circular quantum wire, there are four linear independent backscattering terms, two of them take the value $\theta=0$ and the other two take the value $\theta=\pi/2$. Hence all gapless modes at poles are fully gapped by $\mathcal{L}_{\mathrm{int}}$.

\subsection{$C_4$-symmetric class-$BD\1$ HOTSC}
In the main text, we have defined the $C_4$ symmetry in Eq. ({\color{red}6}) and time-reversal symmetry in Eq. ({\color{red}7}). Now we justify the couplings introduced in the main text that are $(C_4\times\mathbb{Z}_2^T)$-symmetric.

Firstly, it is easy to verify that the generators of 4-fold rotation and time-reversal, $\bs{R}$ and $\mathcal{T}$ satisfy the condition:
\[
\bs{R}^4=\mathcal{T}^2=\mathbbm{1}
\]
We note that the interactions we have introduced in the main text as Eqs. ({\color{red}8})-({\color{red}10}) are $(C_4\times\mathbb{Z}_2^T)$-symmetric only if we define the rotation and time-reversal symmetries as Eqs. ({\color{red}6}) and ({\color{red}7}) in the main text. 

Then we focus on the inter-wire and intra-wire backscattering terms and demonstrate that they can fully gap out the bulk of the $C_4$-symmetric class-$BD\1$ HOTSC. Consider the $(4j-k)^{\mathrm{th}}$ ($k=0,1,2,3$) quantum wires, there are 16 independent bosonic fields $\phi_{1,2}^{4j-k}$. In order to gap them out, we should introduce at least 8 linear independent backscattering terms satisfying the Haldane's ``null-vector'' condition [cf. Eq. (\ref{null-vector})]. It is straightforwardly to verify that the inter-wire coupling $\mathcal{L}_{ck}^j$ includes 8 linear independent backscattering terms satisfying the Haldane's ``null-vector'' condition, hence the related four quantum wires are fully gapped by $\mathcal{L}_{ck}^j$.

For nontrivial HOTSC dominated by $\mathcal{L}_{c4}$, the inner-most quantum wire remains gapless at each pole. The neighborhoods of all four poles includes sixteen segments of bosonic field, $\phi_{1,2}(\theta)$, where $\theta\in(-\epsilon+{\beta\pi}/{2},\epsilon+{\beta\pi}/{2})$, $\beta=0,1,2,3$. The intra-wire coupling $\mathcal{L}_{\mathrm{int}}$ includes only four linear independent backscattering terms globally, nonetheless, if we only focus on the gapless segments at poles of the quantum wire, there are eight linear independent backscattering terms, four of them take the value $\theta=0$, and the other four take the value $\theta=\pi/2$. Hence all gapless modes at poles are fully gapped by $\mathcal{L}_{\mathrm{int}}$.

\providecommand{\noopsort}[1]{}\providecommand{\singleletter}[1]{#1}%

\end{document}